\documentstyle[myltwol]{article}

\input my_psfig.sty

\begin{document}


\newcommand{\cm}{\rm \,cm}
\newcommand{\mm}{\rm \,mm}
\newcommand{\m}{\rm \,m}
\newcommand{\nm}{\rm \,nm}
\newcommand{\minu}{\rm \,min}
\newcommand{\Km}{\rm \,Km}
\newcommand{\e}{\rm e}
\newcommand{\s}{\rm \,s}
\newcommand{\ms}{\rm \,ms}
\newcommand{\sr}{\rm \,sr}
\newcommand{\ns}{\rm \,ns}
\newcommand{\mg}{\rm \,mg}
\newcommand{\lit}{\rm \,l}
\newcommand{\g}{\rm \,g}
\newcommand{\MeV}{\rm \,MeV}
\newcommand{\eV}{\rm \,eV}
\newcommand{\KeV}{\rm \,KeV}
\newcommand{\GeV}{\rm \,GeV}
\newcommand{\TeV}{\rm \,TeV}
\newcommand{\erg}{\rm \,erg}
\newcommand{\Kpc}{\rm \,Kpc}
\newcommand{\Mpc}{\rm \,Mpc}
\newcommand{\kHz}{\rm \,kHz}
\newcommand{\mHz}{\rm \,mHz}
\newcommand{\MHz}{\rm \,MHz}
\newcommand{\Hz}{\rm \,Hz}
\newcommand{\p}{\rm p}
\newcommand{\n}{\rm n}
\newcommand{\deu}{\rm d}
\newcommand{\nue}{$\nu_e$}
\newcommand{\anue}{$\bar{\nu}_e$}
\newcommand{\numu}{$\nu_{\mu}$}
\newcommand{\anumu}{$\bar{\nu}_\mu$}
\newcommand{\nutau}{$\nu_{\tau}$}
\newcommand{\anutau}{$\bar{\nu}_\tau$}
\newcommand{\nux}{$\nu_x$}
\newcommand{\anux}{$\bar{\nu}_x$}
\newcommand{\lsim}{\lower .5ex\hbox{$\buildrel < \over {\sim}$}}
\newcommand{\gsim}{\lower .5ex\hbox{$\buildrel > \over {\sim}$}}
\newcommand{\system}[1]{\left\{\matrix{#1}\right.}
\newcommand{\displayfrac}[2]{\frac{\displaystyle
 #1}{\displaystyle #2}} 
\newcommand{\diff}{{\rm\,d}}

\title{Neutrinos from supernov\ae: experimental status and perspectives \\
\vspace{0.2cm}
{\footnotesize{Invited talk at {\it\lq\lq Matter, Anti-Matter and 
Dark Matter\rq\rq}~ECT$^{\star}$ Workshop, Trento (Italy), 
$29-30$ October $2001$}}
}

\author{Fabrizio Cei}

\address{Dipartimento di Fisica dell' Universit\`a and INFN - Pisa, 
Via Livornese, 1291a, 56018 S. Piero a Grado (PI), \\E-mail: 
fabrizio.cei@pi.infn.it}   

\twocolumn[\maketitle\abstracts{I discuss the state of the art in the 
search for neutrinos from galactic stellar collapses and the future 
perspectives of this field. The implications for the neutrino physics 
of a high statistics supernova neutrino burst detection by the network 
of detectors operating around the world are also reviewed.}]

\section{Introduction}\label{sec:1}
\begin{figure}[htb]
\begin{center}
\mbox{
	\psfig{file=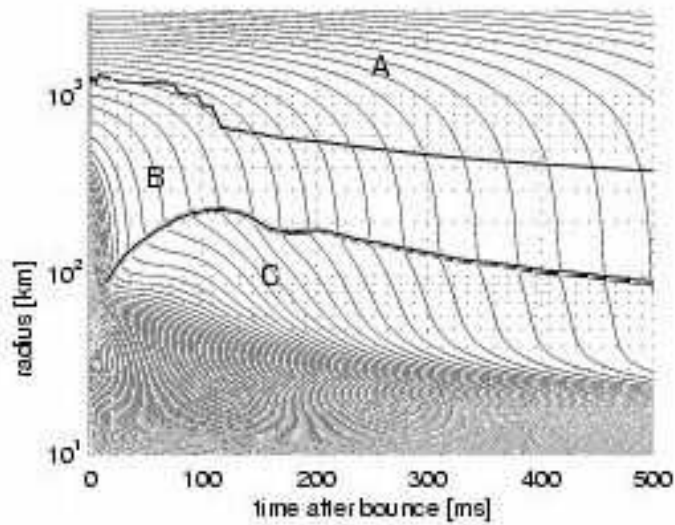,height=3.3in}
     }
\end{center}
\vspace{-1.0cm}
\caption{The trajectories of various mass shells ($0.01~M_{\odot}$), 
the shock (thick line) and the nuclear burning front (thin line) 
in the $\left( time, radius \right)$ plane. A: Silicon; B: Iron produced 
by infall and heating; C: Free nucleons (from [MEZ00b]).}
\label{mez1}
\end{figure}

The core collapse (type II and Ib) supernov\ae~are spectacular events 
which are being studied, by using 
numerical simulations, since more than three decades (see e.g. 
\cite{COL66}, \cite{WIL85}, \cite{BET90}, \cite{BUR90}, \cite{WIL93}, 
\cite{JAN94}, \cite{BUR98}, \cite{JAN96}, \cite{JAN98}, \cite{THO01}, 
\cite{MEZ01}, \cite{RAF01}). Despite the huge amount of physics involved 
in these catastrophic explosions, a sort of \lq\lq supernova 
standard model\rq\rq~has been emerging in the last years 
(\cite{MES98}, \cite{MEZ00b}): the inner Iron 
core of a massive star ($M~\gsim \; 8~M_{\odot}$) overcomes 
its hydrodinamical stability limit (the Chandrasekhar mass) and collapses, 
raising its density up to many times the nuclear density; this anomalous 
density produces an elastic bounce of the core, which results in a shock 
wave. The wave propagates through the star, loses energy in dissociating 
nucleons and producing neutrinos and finally stalls at $\sim 200~{\rm Km}$ 
from the center of the star (this means that the so-called \lq\lq 
prompt\rq\rq~mechanism unavoidably fails). However, {\nue} and {\anue} 
are absorbed on the nucleons liberated by the shock; such processes 
supply new energy to the wave, which is revived, $\sim 500~{\rm ms}$ 
after the bounce (this energy transfer is known as \lq\lq 
neutrino heating\rq\rq). The reinforced shock can propagate within 
the stellar matter and expel the external layers into the space 
(this is the so-called \lq\lq delayed\rq\rq~mechanism). As an 
example of the results of a numerical simulation, Fig. \ref{mez1} 
shows the trajectories of equal mass shells ($0.01~M_{\odot}$), 
the shock and the nuclear burning front in the $\left( time, 
radius \right)$ plane in a $13~M_{\odot}$ model (from \cite{MEZ00b}). 
After the explosion the star loses energy, mainly by neutrino emission, 
and cools down, forming a neutron star or a black hole. 

In this general picture there are, however, still many question marks.  
For instance, is the neutrino heating enough to trigger the 
explosion or some other energy transfer mechanism is needed ? Do the 
convection of the stellar matter, the star rotation and/or the star magnetic 
field play important roles in enhancing the energy transfer ? Is the behaviour 
of the nuclear matter at a density $\sim 10^{14}~{\rm g} \, {\rm cm}^{-3}$ 
really well understood and the neutrino transport inside such a 
highly degenerate matter treated with sufficient accuracy ? Is the 
spherical symmetry of the explosion largely violated, as the 
detection of polarized light and boosted ${ }^{56}Ni$ nuclei in 
the SN$1987$A remnants \cite{WHE00} seems to indicate ? Is the residual 
of the explosion a neutron star or a black hole ? Is there a sort 
of \lq\lq threshold progenitor mass\rq\rq, which 
separates these two different destinies ? 

These are only few of the various open questions which could be 
addressed, at least partially, by the observation of a neutrino burst 
from a galactic supernova (\cite{PRA01}, \cite{TOT98a}). Many features 
of the collapse mechanism are indeed imprinted in the neutrinos 
released during the explosion. At the same time, a galactic supernova 
explosion would give particle physicists the opportunity to explore  
the neutrino properties on scales of distance up to 
$\sim 10^{17}~{\rm Km}$, of time up to $\sim 10^{5}~{\rm years}$ and 
of density beyond that of the nuclear matter.
 
Here I will focus on the real time detection of supernova neutrino 
bursts, without considering the ideas and techniques developed to search 
for relic neutrinos from past supernov\ae.   
\section{Supernova Neutrino Bursts}\label{sec:2}
A supernova explosion releases $\sim 2 \div 4 \times 10^{53}~{\rm ergs}$ of 
gravitational binding energy. The kinetic energy of the expelled matter is 
lower by about two orders of magnitude and the energy emitted in 
electromagnetic radiation and gravitational waves is even less; then, the 
bulk of the energy is released in the form of neutrinos. 

A first {\nue} burst is emitted during the infall stage of 
the collapse ({\it\lq\lq infall burst\rq\rq}), since the high density of 
matter makes the electron capture by proton: 
\begin{equation}
e^{-} + p \rightarrow \nu_e + n \\
\label{ep}
\end{equation}
very efficient. However, this neutrino emission does not continue 
indefinitely, since at a density $\sim 10^{12}~{\rm g \, cm^{-3}}$ the 
neutrinos are trapped in the stellar core and go into equilibrium 
with matter via the inverse process:  
\begin{equation}
\nu_e + n \rightarrow e^{-} + p  
\label{nuen}
\end{equation}
Immediately after the core bounce, also neutrinos of other flavours begin 
to be produced, via nucleonic bremsstrahlung and pair annihilation 
processes as:
\begin{equation}
e^{+} + e^{-} \rightarrow \nu_{x} + \bar{\nu}_{x}
\label{anni}
\end{equation}
The neutrinos are trapped in a region (the {\it\lq\lq 
neutrinosphere\rq\rq}) whose size is different for different 
neutrino flavours. Neutrinos produced at a distance from the 
center larger than the neutrinosphere radius can freely escape to infinity, 
while neutrinos produced within this sphere remain trapped. 
Electron neutrinos and antineutrinos interact with matter via 
neutral and charged current processes (as (\ref{nuen}) or 
$\bar{\nu}_{e} + p \rightarrow e^{+} + n$), while non-electron 
neutrinos and anti-neutrinos (from now on, collectively indicated with 
\lq\lq {\nux}\rq\rq) interact only via neutral currents. Therefore, 
the {\nux} are less tightly coupled with matter than {\nue} and 
{\anue} and a higher matter density is needed to trap them. Moreover, 
since the mantel is richer in neutrons than in protons (because of 
the (\ref{ep}) process), the {\nue} are more strongly coupled than 
the {\anue}. As a result, the neutrinosphere radius is maximum for 
{\nue} and amounts to $\sim 70~{\rm Km}$ for {\nue}, to 
$\sim 50~{\rm Km}$ for {\anue} and to $\sim 30~{\rm Km}$ for {\nux}.
The larger the neutrinosphere radius, the lower the mean neutrino 
energy (a deeper neutrinosphere corresponds to a higher temperature); 
so, the neutrino mean energies are expected to be (see e.g. \cite{JAN93}):
%
\begin{eqnarray}
\langle E_{\nu_{e}} \rangle \approx 10 \div 13~{\rm MeV} \\
\label{avenue}
\langle E_{\bar{\nu}_{e}} \rangle \approx 14 \div 17~{\rm MeV} \\
\label{aveanue}
\langle E_{\nu_{x}} \rangle \approx 22 \div 27~{\rm MeV} 
\label{avenux}
\end{eqnarray}
When the shock crosses the {\nue} neutrinosphere, an intense burst of 
{\nue} is produced, since the efficiency of the (\ref{ep}) process is 
abruptly enhanced by the large number of protons liberated by the 
shock ({\it\lq\lq neutronization burst\rq\rq}). The infall and 
neutronization bursts are very rapid ($\sim 10~{\rm ms}$); the {\nue} 
energy is $\sim 10~{\rm MeV}$ and the total released energy is 
$\lsim~10^{52}~{\rm erg}$. 
 
After the neutronization stage, the {\nue} luminosity rapidly decreases, 
while the luminosities of other flavour neutrinos increase. The neutrinos 
are now produced mainly via the (\ref{anni}) process and the higher rate 
of {\nue}, {\anue} production is compensated by the larger coupling of 
these neutrinos with matter. As a result, at the end of the neutrino 
diffusion inside the mantle the energy is practically equally distributed 
between the various neutrino flavours ($\sim 5 \times 10^{52}~{\rm ergs}$ 
for each neutrino type). This final stage ({\it\lq\lq cooling\rq\rq}) 
requires $\sim 10~{\rm s}$ and takes away $> 99 \, \%$ of the 
gravitational energy of the star. 

\begin{figure}[htb]
\begin{center}
\mbox{
	\psfig{file=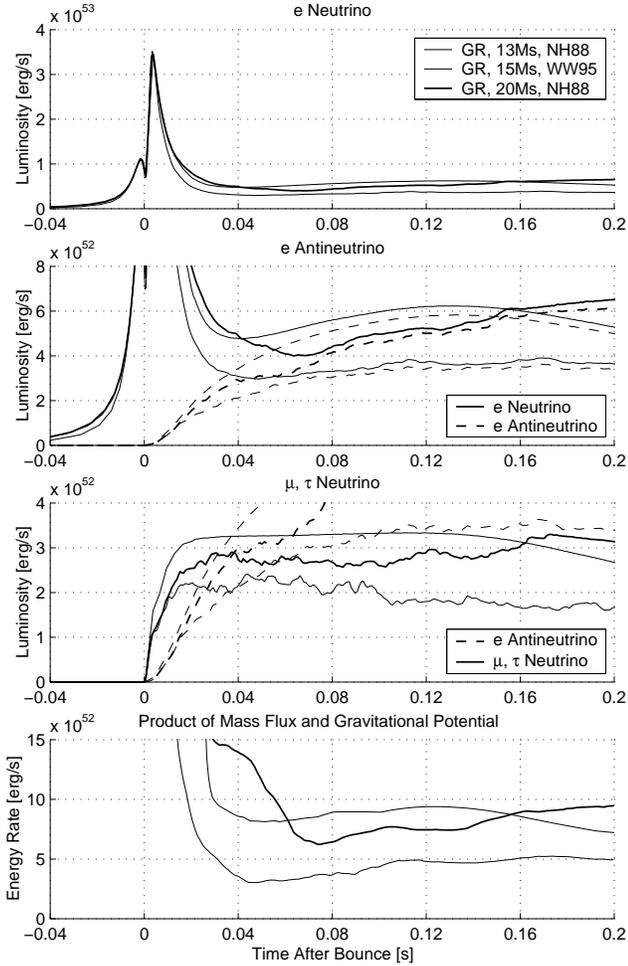,height=5.0in}
     }
\end{center}
\vspace{0.7cm}
\caption{Luminosities for all kind of neutrinos as a function of time 
after the bounce (first three graphs). In any graph three different 
curves are shown, each one corresponding to a different explosion model.
The last graph shows the energy production rate, closely similar to 
the neutrino luminosity (from [LIE01]).}
\label{mez}
\end{figure}

Fig. \ref{mez} shows an example of the neutrino luminosities 
as a function of time after the bounce \cite{LIE01}. 
Note the {\nue} neutronization peak, the sharp rise (tenths of {\rm ms}) 
of all flavour luminosities and the much longer trailing edge 
of all neutrino signals. 

\begin{figure}[htb]
\begin{center}
\mbox{
	\hspace{-0.3cm}\psfig{file=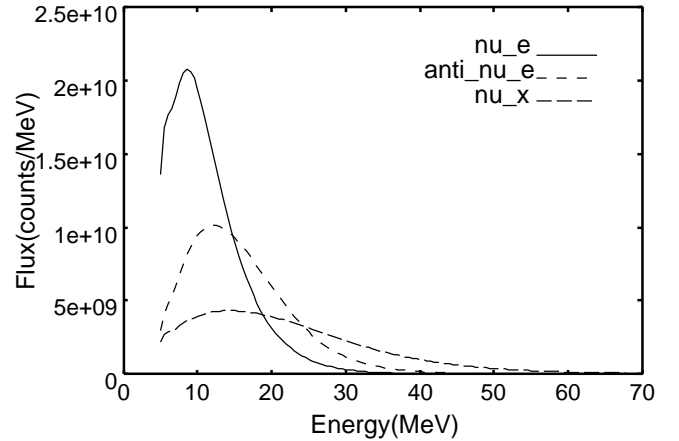,height=2.6in}
     }
\end{center}
\caption{Neutrino and antineutrino energy spectra; {\nux}  
indicates non-electron neutrinos or antineutrinos (from [TOT98a]).}
\label{totani}
\end{figure}

The neutrino energy spectra at the time of decoupling reflect the fact 
that, inside the core, they are in equilibrium with matter; so, thermal 
(Fermi-Dirac) or quasi-thermal spectra are obtained by numerical simulations; 
an example is shown in Fig. \ref{totani} (from \cite{TOT98a}). 

The general features of neutrino spectra and luminosities are rather 
similar in the various simulations; some of them (average {\anue} energy, 
duration of the burst, energy emitted in {\anue} etc.) were also confirmed 
by the observation of SN$1987$A neutrinos. In the rest of this paper I 
will refer to the \cite{BUR92} paper as a benchmark for computing the 
expected neutrino signals in various detectors. Other models 
predict harder spectra (expecially for {\nux}), so that the expected 
fractions of the various types of neutrino events can change by many percent, 
mainly for the {\nux}-induced neutral current reactions. 

\section{Supernova neutrino reactions}\label{sec:3}
The detection of supernova neutrinos requires their conversion into a 
charged lepton or the emission of some other particles (essentially 
$\gamma$'s or neutrons) which can be efficiently observed. An \lq\lq 
ideal\rq\rq~reaction should be sensitive to all flavours of neutrinos 
and preserve all information on neutrino energy spectrum, timing, direction 
and type; however, usually only a part of this information can be derived 
by a single process and then a combined observation by detectors based 
on different technologies is fundamental to obtain a comprehensive picture 
of the neutrino burst properties. 

\begin{figure}[htb]
\begin{center}
\mbox{
	\hspace{-0.2cm}\psfig{file=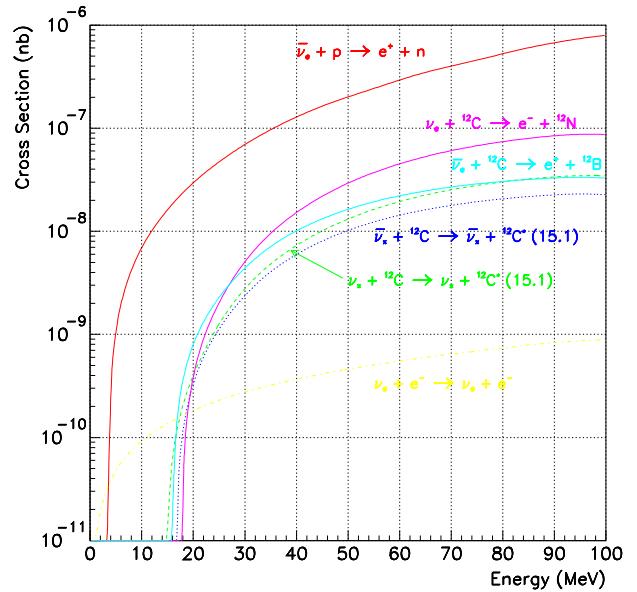,height=3.6in}
     }
\end{center}
\caption{Cross sections for CCp, ES, CC and NC on Carbon processes.}
\label{nucros}
\end{figure}

\begin{figure}[htb]
\begin{center}
\mbox{
        \hspace{-0.2cm}\psfig{file=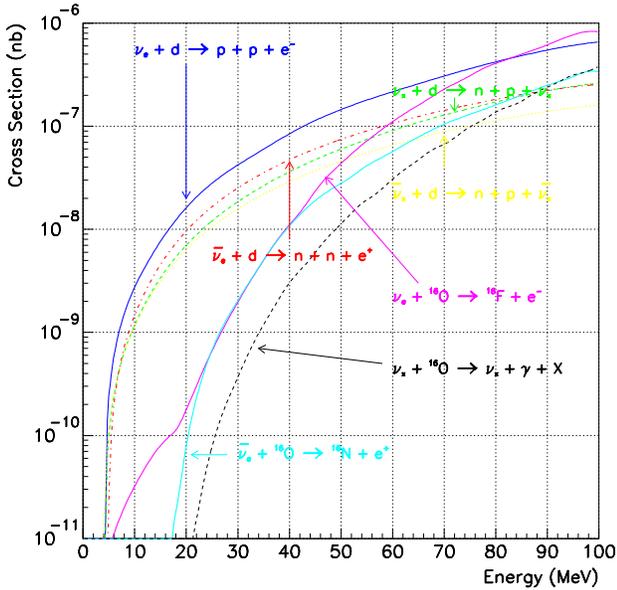,height=3.6in}
     }
\end{center}
\caption{Cross sections for neutrino reactions on Deuterium and 
Oxygen. The shoulder at low energies of the $\nu_e$-Oxygen cross 
section comes from the contribution of ${ }^{18}O$ ($< 0.1 \, \%$ of 
isotopic composition of the normal Oxygen).}
\label{nucrosdeu}
\end{figure}

Here I briefly review the reactions employed in the supernova neutrino 
detection, which fall in four main cathegories. 
\begin{itemize}
\item {\bf Charged currents on nucleons (CCn(p)).} These processes are 
useful only for {\nue} and {\anue}, because the {\nux} energies are 
under the threshold for charged leptons 
($\mu$ and $\tau$) production. Target materials containing free neutrons 
are not available; then, one is left with the inverse-beta reaction:
\begin{equation}
\bar{\nu}_e + p \rightarrow e^{+} + n
\label{anuep}
\end{equation}
This reaction is the most interesting one in proton-rich targets, like water 
or hydrocarbonic scintillators. Its cross section is proportional to the 
square of the neutrino energy, but it is usually written in the following 
form \cite{BAH89}: 
\begin{equation}
\sigma = 8.5 \times 10^{-44} ~ \left( E_{e^{+}} \left( {\rm MeV} \right) 
\right)^2~{\rm cm}^2
\label{cranuep}
\end{equation}
which contains the directly measurable $e^{+}$ energy, related to the 
{\anue} energy by the formula $E_{e^{+}} \approx E_{\bar{\nu}_e} - 
m_n + m_p = E_{\bar{\nu}_e} - 1.293~{\rm MeV}$. A further 
$1.022~{\rm MeV}$ energy can be detected because the positron 
annihilates with an electron, producing a pair of $0.511~{\rm MeV}$ 
photons. The neutron emitted in the process (\ref{anuep}) can be 
thermalized and captured by a proton, forming a Deuterium nucleus 
and releasing the binding energy as a $2.2~{\rm MeV}$ $\gamma$ (from now 
on, $\gamma_{2.2}$). In liquid scintillators the average 
moderation time is $\sim 10~{\rm \mu s}$ and the average capture time 
is $\sim 180~{\rm \mu s}$. The detection of $\gamma_{2.2}$ is a 
further signature of the (\ref{anuep}) reaction and can be an important 
tool to separate the (\ref{anuep}) positrons from the products of 
other reactions. The threshold for the process (\ref{anuep}) is 
$1.8~{\rm MeV}$, a value which cuts off only a small fraction 
($\lsim \; 5 \, \%$) of the {\anue} spectrum. All SN$1987$A neutrinos 
observed by Kamiokande II \cite{KAM87}, IMB$3$ \cite{IMB87} and 
Baksan \cite{BAK87} are generally believed to be {\anue}, detected 
via the (\ref{anuep}) process. The reaction (\ref{anuep}) has only a 
weak dependence on the neutrino incoming direction \cite{VOG99}: 
the average value of the cosine of the angle between {\anue} and 
$e^{+}$ varies from $\approx -0.03$ to $\approx 0.1$ 
in the supernova neutrino energy range.

\item {\bf Elastic scattering on electrons (ES).} This reaction is possible, 
in principle, for all kind of neutrinos; however, {\nue} and {\anue} interact 
with electrons via charged (with different couplings) and neutral currents, 
while {\nux} interact only via neutral currents. The {\nue} have the  
highest cross section: 
\begin{equation}
\sigma_{\nu_{e} \,e} = 9.2 \times 10^{-45}~E_{\nu_{e}} 
( {\rm MeV} )~{\rm cm}^2.
\label{esnue}
\end{equation} 
and the following cross section hierarchy holds:
\begin{equation}
\sigma_{\nu_{e} \, e} \approx 3 \,\sigma_{\bar{\nu}_{e} \, e} 
\approx 6 \,\sigma_{\nu_{x} \, e}
\label{esall}
\end{equation}
The number of ES events expected in a detector is much lower than that 
expected for (\ref{anuep}) reactions because of the lower multiplicative 
coefficient and of the linear dependence of the cross section 
(\ref{esnue}) on the neutrino energy. However, the angular distribution 
of the ES electrons is strongly peaked around the neutrino incoming 
direction, with an opening angle 
$\theta_{\nu_{e},e}  \sim \left( m_e/E_{\nu_{e}} \right)^{1/2}$ \cite{BAH89}; 
the ES is then a useful reaction for determining the supernova direction. The 
Kamiokande \cite{KAM96}, Super-Kamiokande \cite{BLA01} and SNO 
\cite{WAL01} experiments detected thousands of ES events induced by solar 
neutrinos. Note that the ratio between the number of (\ref{anuep}) and 
(\ref{esnue}) events is sensitive to the neutrino energy spectra and is 
then an important tool in discriminating between various supernova 
models \cite{SCH88}.

\item {\bf Charged currents on nuclei (CCN).} Again, only {\nue} and {\anue} 
can interact via charged current reactions. A general feature of CCN 
reactions is a relatively high cross section, in some cases competitive 
with that of the CCn reactions for neutrino energies 
$\gsim~20 \div 30~{\rm MeV}$. These reactions are followed by the 
$\beta^{\pm}$-decay ($\tau \sim 20~{\rm ms}$) of the products nuclei, so 
that they have, in principle, a very clean signature.  
However, their practical application is usually strongly 
limited by the high threshold ($\sim 15~{\rm MeV}$ for interactions with 
$C$ or $O$ nuclei), which cuts off the most part of the {\nue}, {\anue} 
spectrum. 

A very important exception is the Deuterium, which can be disintegrated 
by {\nue}, {\anue} via the processes (for recent evaluations of the 
cross sections see \cite{BUT01}, \cite{NAK01}):
\begin{eqnarray}  
      \nu_e + d \rightarrow e^{-} + p + p 
\label{nued} \\
\bar{\nu}_e + d \rightarrow e^{+} + n + n 
\label{anued}
\end{eqnarray}
The thresholds for reactions (\ref{nued}) and (\ref{anued}) are respectively 
$1.44~{\rm MeV}$ and $4.03~{\rm MeV}$; both these values are well below the 
average {\nue}, {\anue} energy. The angular distributions of both the 
processes have the form $d\sigma/d\Omega \propto 1 - a \left( E_{\nu} \right) 
\,\cos \left( \theta \right)$, where $\theta$ is the angle between {\nue} 
({\anue}) and $e^{-}$ ($e^{+}$) and $a \left( E_{\nu} \right)$ is an 
energy-dependent coefficient. In the supernova neutrino energy range, 
$a \left( E_{\nu} \right)$ is nearly costant at $\approx 1/3$ for {\nue}, 
while for {\anue} it is proportional to $E_{\nu}$ and ranges from 
$\approx 1/3$ at $E_{\nu} = 0$ to $\approx 0$ at $E_{\nu} = 50~{\rm MeV}$. 
The recent measurement of the solar neutrino flux by the 
SNO experiment \cite{WAL01} via the (\ref{nued}) process showed that such  
reaction can be succesfully employed for detecting low energy neutrinos 
and that its contribution can be statistically separated by that due to 
ES reactions.

An other interesting CCN reaction is:
\begin{equation}
\nu_{e} ~+~ { }^{40}Ar \rightarrow { }^{40}K^{*} ~+~e^{-} 
\label{ArLi}
\end{equation}
which can take place in Liquid Argon detectors \cite{ORM95}. 
This reaction has a $5.885~{\rm MeV}$ threshold and is accompanied by the 
${ }^{40}K^{*}$ de-excitation to the ground state, which releases a 
$5~{\rm MeV}$ $\gamma$. The cross section of this reaction overcomes 
that of the (\ref{anuep}) process for $E_{\nu_{e}} \gsim~20~{\rm MeV}$.  
\item {\bf Neutral currents on nuclei (NC).}      
As one can argue from the previous discussion, the detection of {\nux}  
is the most challenging goal of the supernova neutrino detectors. 
Despite the larger difficulties, the detection of {\nux} is extremely 
important, not only for astrophysics, but also for particle physics, because 
of the opportunity to set stringent limits on the masses of non-electronic 
neutrinos (this point will be examined in section \ref{sec:54}). Two kinds of 
possible processes were suggested: the excitation and subsequent 
de-excitation of a nuclear level, accompanied by the emission of a photon, 
and the knocking off of a neutron. Note that none of these reactions is 
sensitive to the neutrino energy or direction, since the energy of the 
de-excitation photon is set by the nuclear level, the neutron recoil can 
not be measured and the neutron and photon emissions are essentially 
isotropic. However, the neutrino arrival time can be measured. 

Neutrinos can excite ${ }^{12}C$ to a $15.1~{\rm MeV}$ level (the 
de-excitation photon will be indicated, from now on, by $\gamma_{15.1}$). 
The cross sections for neutrino-Carbon reactions were measured by the 
LAMPF \cite{LAM92}, Karmen \cite{KAR93} and LSND \cite{LSN97} collaborations 
and the experimental values were in agreement with theoretical calculations 
\cite{FUK88}. Because of the high threshold ($15.1~{\rm MeV}$), this NC 
process is a good selector of {\nux}, whose spectrum is harder 
than that of {\nue}, {\anue}. 

Neutrinos can also excite ${ }^{16}O$ via various anelastic processes 
with emissions of protons (or neutrons) and $\gamma$'s, with a total 
photon energy between $5$ and $10~{\rm MeV}$ \cite{LAN96}. As for 
reactions on Carbon, only the more energetic {\nux} can efficiently 
excite ${ }^{16}O$ nuclei because of the high energy threshold.

The Deuterium nucleus can be disintegrated by neutrinos of all flavours 
via the NC process:
\begin{equation}
\nu_{x} + d \rightarrow \nu_{x} + n + p
\label{nudisin}  
\end{equation}
The cross section for reaction (\ref{nudisin}) was computed by many 
authors (e.g. \cite{BUT01}, \cite{NAK01}) and is competitive with  
that of (\ref{anuep}) process; the threshold is $2.2~{\rm MeV}$, so that 
also {\nue} and  {\anue} can efficiently break-up the Deuterium nuclei. 
(The process (\ref{nudisin}) can indeed give a flavour-independent 
measurement of the solar neutrino flux and was originally suggested having 
such a goal in mind.) 

Finally, during the last years some theoretical calculations (e.g. 
\cite{CLI94}) pointed out that the cross sections for neutron knocking off 
from certain heavy nuclei ($Ca$, $Na$, $Pb$ ...) are large 
($\sim 10^{-42}~{\rm cm}^2$) and steeply increasing at energies 
$\gsim~20~{\rm MeV}$, so that the more energetic {\nux} could be 
efficiently selected. The use of chemical compounds of high solubility 
in water, as $Pb \left(ClO_{4} \right)_{2}$, has also been suggested 
to combine the high value of the CCN and NC neutrino cross sections 
on high-Z nuclei with the experimental advantages offered by 
the water \v{C}erenkov technology \cite{ELI00}. By now these cross section 
calculations have no experimental support, but proposed new facilities, 
like {\bf ORLaND} at Oak Ridge National Laboratory \cite{MEZ00a}, 
should provide some measurements of neutrino-nucleus cross sections 
of interest in supernova neutrino detection.    
\end{itemize}
To summarize, Fig. \ref{nucros} shows the cross sections (in {\rm nb}) 
for CCp, ES, CC and NC on Carbon processes and Fig. \ref{nucrosdeu} the 
cross sections for the various neutrino reactions on Deuterium 
and Oxygen.  

\section{Present supernova neutrino detectors}\label{sec:4}
\subsection{General considerations}\label{sec:41}
The supernov\ae~are very rare events: the expected rate of galactic stellar 
collapses is $\sim 1/20 \div 40~{\rm years}$ (\cite{VAN93}, \cite{TAM94}) 
(note that this is a, so to say, \lq\lq unlucky\rq\rq~number: it is too 
low to give a good chance of success to an experiment with an operating 
life of $\sim 10$ years and, at the same time, is too high to make a 
real time detection of a supernova neutrino burst a hopeless dream). 
Moreover, the supernova neutrinos have energies of tenths of MeV or 
less, with low cross sections (as discussed above) and, finally, 
a supernova has (at least can have\footnote{The optical flare can 
be absent because the explosion \lq\lq fizzles\rq\rq~or because the 
supernova is in a sky region optically obscured by the halo luminosity 
or the cosmic dust.}) a bright optical flare and releases radiation 
in other forms than neutrinos. These three points set the fundamental 
guidelines in designing a good supernova neutrino detector:
\begin{itemize}
\item[$1)$] the detector must have a mass $\gsim~10^3~{\rm tonn}$ of active 
material. Such material must be cheap, robust and possibly not polluting;  
\vspace{-0.2cm}
\item[$2)$] the detector must be located underground, to reduce the cosmic 
ray induced background, and possibly in a low radioactivity environment;
\vspace{-0.2cm}
\item[$3)$] the detector must have a very high duty cycle (in principle it 
should be always active) and an operating life of at least 
$\sim 10~{\rm years}$;  
\vspace{-0.2cm}
\item[$4)$] the detector must be equipped with electronics and acquisition 
systems well suited to perform a real time neutrino detection, with 
good accuracies in absolute ($\lsim~1~{\rm ms}$) and relative 
($\lsim~1~\mu{\rm s}$) timing and (if possible) good angular and energy 
resolutions. The energy threshold should not exceed $\sim 10~{\rm MeV}$. 
\end{itemize}
I now review the presently active neutrino detectors and discuss how (and 
whether) the requests listed above are satisfied. Note that most of these 
detectors were build having in mind also other physics goals, as the detection 
of solar and atmospheric neutrinos, the search for magnetic monopoles 
and proton decay, the observation of high energy cosmic rays etc. 
\subsection{Scintillation detectors}\label{sec:42}
Scintillation detectors use large masses ($\sim 10^{3}~{\rm tonn}$) of high 
transparency mineral oils ($CH_{n}$, $n = 1,2$), segmented in hundreds of 
individual counters or enclosed in a container and observed by hundreds of 
PMTs at the boundary of the active volume. The scintillation detectors are 
mainly sensitive to {\anue} via the process (\ref{anuep}); a contribution of 
$3 \div 6 \, \%$ of the total number of events is expected from the 
NC on Carbon reactions and even lower contributions from ES and CC on Carbon 
processes. The scintillation detectors have a larger light yield and a 
better energy resolution than the water \v{C}erenkov detectors; then, 
they are sensitive to the $\gamma_{2.2}$. 

\begin{figure}[htb]
\begin{center}
\mbox{
	\psfig{file=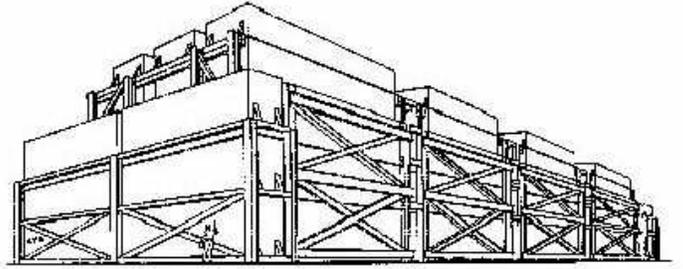,width=3.8in}
     }
\end{center}
\caption{The LVD experiment.}
\label{LVD}
\end{figure}

The {\bf LVD} \cite{LVD92} ({\bf L}arge {\bf V}olume {\bf D}etector) 
experiment, shown in Fig. \ref{LVD}, is presently the largest operating 
liquid scintillation detector in the world.
This detector is located in the Hall \lq\lq A\rq\rq~of the Laboratory 
Nazionali del Gran Sasso (LNGS), in the central Italy, at an average depth 
of $3100~{\rm m.w.e.}$. After many years of operation with $\sim 
600~{\rm tonn}$ of liquid scintillator, LVD reached this year 
its final active mass of $\approx 10^{3}~{\rm tonn}$, organized in $840$ 
counters ($1 \times 1 \times 1.5~{\rm m}^3$ each) interleaved with Streamer 
Tubes for cosmic muons tracking \cite{VIG01}. This structure is 
partially self-shielded, with the internal counters ($\sim 30 \, \%$ of 
the total) subject to a lower radioactive background than the external ones. 

The {\bf MACRO} (\cite{MAC92}, \cite{MAC98a}) ({\bf M}onopole 
{\bf A}strophysics and {\bf C}osmic 
{\bf R}ay {\bf O}bservatory) experiment, also located in the LNGS 
(Hall \lq\lq B\rq\rq) and recently ended (December $2000$), with 
$570~{\rm tonn}$ of active mass had a sensitivity similar to that of LVD. 
This detector was equipped with $476$ very long counters ($\approx 
12~{\rm m}$ each), deployed in horizontal and vertical layers. 

Table \ref{scinev} compares the LVD and MACRO properties 
in neutrinos from stellar collapse detection. The upper part of the 
table shows the expected number of events in these two detectors 
from a supernova at the Galactic Center (distance $8.5~{\rm Kpc}$) 
and the lower part compares their performances (resolution, energy 
threshold $E_{thr}$ etc.) at supernova neutrino energies.
\begin{table}[htb]
\begin{center}
\caption{Upper part: expected number of events in LVD ($10^{3}~{\rm 
tonn}$) and MACRO from a supernova at the Galactic Center. Lower part: 
comparison between the LVD and MACRO performances at $E = 10~{\rm MeV}$. 
(\lq\lq E\rq\rq~and \lq\lq I\rq\rq~indicate the external and internal 
LVD counters.)}
\label{scinev}
\vspace{0.2cm}
\begin{tabular}{|c|c|c|c|c|} 
\hline
\hline 
\multicolumn{5}{|c|} {\bf Number of expected events} \\ \cline{1-5}
\hline
\raisebox{0pt}[12pt][6pt]{ } & \multicolumn{2}{c|} {\bf LVD} & 
                               \multicolumn{2}{c|} {\bf MACRO} \\ \cline{2-5}
\hline
\raisebox{0pt}[12pt][6pt]{ } &
\raisebox{0pt}[12pt][6pt]{\bf Events} &
\raisebox{0pt}[12pt][6pt]{\bf (\%)} &
\raisebox{0pt}[12pt][6pt]{\bf Events} &
\raisebox{0pt}[12pt][6pt]{\bf (\%)} \\
\hline
\raisebox{0pt}[12pt][6pt]{$\bar{\nu}_{e}+ p$} & 
\raisebox{0pt}[12pt][6pt]{$296$} & {$93.0$} &
\raisebox{0pt}[12pt][6pt]{$198$} & {$94.7$} \\
\raisebox{0pt}[12pt][6pt]{${\nu}_{e} + e$} & 
\raisebox{0pt}[12pt][6pt]{$5$} & {$1.6$} &
\raisebox{0pt}[12pt][6pt]{$3$} & {$1.4$} \\ 
\raisebox{0pt}[12pt][6pt]{$\bar{\nu}_{e} + e$} & 
\raisebox{0pt}[12pt][6pt]{$1$} & {$0.3$} &
\raisebox{0pt}[12pt][6pt]{$< 1$} & {$<0.5$} \\ 
\raisebox{0pt}[12pt][6pt]{${\nu}_{x} + e$} & 
\raisebox{0pt}[12pt][6pt]{$3$} & {$0.95$} &
\raisebox{0pt}[12pt][6pt]{$2$} & {$0.96$} \\ 
\raisebox{0pt}[12pt][6pt]{$\nu_{e} + C$ ({\bf CC)}} & 
\raisebox{0pt}[12pt][6pt]{$\approx 1$} & {$0.3$}& 
\raisebox{0pt}[12pt][6pt]{$< 1$} & {$< 0.5$} \\ 
\raisebox{0pt}[12pt][6pt]{$\bar{\nu}_{e} + C$ ({\bf CC)}} & 
\raisebox{0pt}[12pt][6pt]{$\approx 1$} & {$0.3$} &
\raisebox{0pt}[12pt][6pt]{$< 1$} & {$<0.5$} \\ 
\raisebox{0pt}[12pt][6pt]{$\nu_{x} + C$ ({\bf NC)}} & 
\raisebox{0pt}[12pt][6pt]{$11$} & {$3.5$} &
\raisebox{0pt}[12pt][6pt]{$4$} & {$1.9$} \\ 
\hline
\raisebox{0pt}[12pt][6pt]{\bf Total on p} &
\raisebox{0pt}[12pt][6pt]{$296$} & {$93.0$} &
\raisebox{0pt}[12pt][6pt]{$198$} & {$94.7$} \\
\raisebox{0pt}[12pt][6pt]{\bf Total on e} &
\raisebox{0pt}[12pt][6pt]{$9$} & {$2.8$} &
\raisebox{0pt}[12pt][6pt]{$5 \div 6$} & {$2.4 \div 2.9$} \\
\raisebox{0pt}[12pt][6pt]{\bf Total on C} &
\raisebox{0pt}[12pt][6pt]{$13$} & {$4.1$} &
\raisebox{0pt}[12pt][6pt]{$5 \div 6$} & {$2.4 \div 2.9$} \\
\hline
\raisebox{0pt}[12pt][6pt]{\bf Total} &
\raisebox{0pt}[12pt][6pt]{$318$} & {$100$} &
\raisebox{0pt}[12pt][6pt]{$209$} & {$100$} \\
\hline
\hline
\multicolumn{5}{|c|} {\bf Performances} \\ \cline{1-5}
\hline
\raisebox{0pt}[12pt][6pt]{ } & \multicolumn{2}{c|}{\bf LVD}  & 
                               \multicolumn{2}{c|}{\bf MACRO} \\ \cline{2-5}
\hline
\raisebox{0pt}[12pt][6pt]{\bf $\boldmath{E_{thr}}$ (MeV)} &  
                     \multicolumn{2}{|c|}{$7$ (E), $4$ (I)} &
                     \multicolumn{2}{|c|}{$7$} \\ 
\raisebox{0pt}[12pt][6pt]{$\frac{\sigma_E}{E} \, (\%)$} & 
                         \multicolumn{2}{|c|}{$15$} &
                         \multicolumn{2}{|c|}{$10$} \\ 
\raisebox{0pt}[12pt][6pt]{$\sigma_t \, ({\rm ns})$} & 
                         \multicolumn{2}{|c|}{$12.5$} &
                         \multicolumn{2}{|c|}{$1$} \\ 
\raisebox{0pt}[12pt][6pt]{$\epsilon_{\gamma_{2.2}} \, (\%)$} &
                         \multicolumn{2}{|c|}{$70$} &
                         \multicolumn{2}{|c|}{$25$} \\
\raisebox{0pt}[12pt][6pt]{$\epsilon_{\gamma_{15.1}} \, (\%)$} &
                         \multicolumn{2}{|c|}{$50$} &
                         \multicolumn{2}{|c|}{$30$} \\ 
\hline
\hline
\end{tabular}
\end{center}
\end{table}

A third scintillation detector, the {\bf Baksan} observatory \cite{BAK98}, 
equipped with $200~{\rm tonn}$ (fiducial volume) of liquid scintillator, 
has been searching for galactic stellar collapses since more than 
$20~{\rm years}$. During $1987$ this detector recorded a $5$ event burst 
which was generally attributed to neutrinos from SN$1987$A \cite{BAK87}.   

Two liquid scintillation detectors of the \lq\lq egg-container\rq\rq~type 
should go on-line soon, {\bf Borexino} \cite{CAD00} in the LNGS 
(Hall \lq\lq C\rq\rq) and {\bf Kamland} \cite{SVO01} in the Kamioka mine 
(Japan). Borexino has a multi-shielding structure, with an internal 
fiducial volume of $300~{\rm tonn}$ of liquid scintillator; Kamland has 
an active volume of $\approx 1000~{\rm tonn}$, surrounded by an 
external buffer of mineral oil and liquid scintillator, used as veto 
for radioactivity and cosmic ray muons. Both these detectors were 
designed to perform a real time detection of very low energy neutrinos 
(solar neutrinos for Borexino and reactor antineutrinos for Kamland) 
and are then characterized by an extremely low radioactivity background 
(the ${ }^{238}U$ and ${ }^{232}Th$ contaminations are at the level 
of $10^{-16}~{\rm g}/{\rm g}$). An other interesting feature of these 
detectors is that, having a large homogeneous active volume, they 
are well suited to detect high energy photons, particularly 
$\gamma_{15.1}$. 
\subsection{Water \v{C}erenkov detectors}\label{sec:43}
Water \v{C}erenkov detectors use large volumes of highly purified water, 
equipped with an array of inward-looking PMTs to detect the \v{C}erenkov 
light produced by relativistic charged particles. The energy and direction 
of the particles can be inferred by the total amount of collected light 
and by the pattern of illuminated phototubes. The \v{C}erenkov detectors 
have a continuous active volume and are self-shielded, i.e. the inner part 
of the detector is shielded from the external radioactivity background by 
the outer one; a fiducial volume can then be defined. The water \v{C}erenkov 
experiments are mainly sensitive to the (\ref{anuep}) process, with few 
per cent contributions from other reactions. The Kamiokande II 
and IMB$3$ detectors, which recorded the well established burst of neutrinos 
from SN$1987$A, were both water \v{C}erenkov experiments.

\begin{figure}[htb]
\begin{center}
\mbox{
	\psfig{file=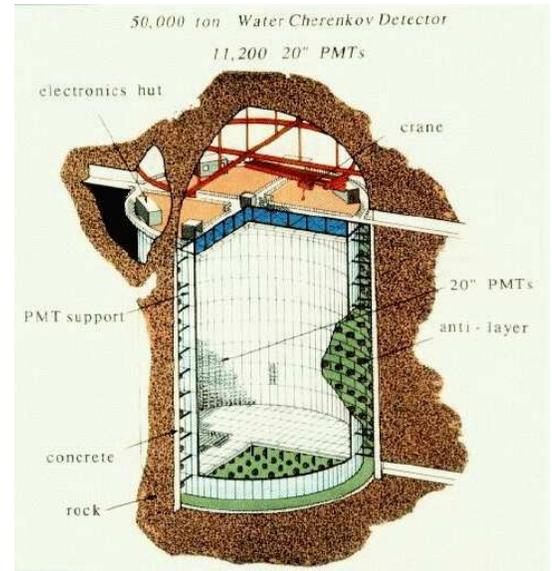,width=2.8in}
     }
\end{center}
\caption{The Super-Kamiokande experiment.}
\label{SK}
\end{figure}

Figure (\ref{SK}) shows the {\bf Super-Kamiokande} detector \cite{SKA98a}, 
a $50~{\rm kton}$ water \v{C}erenkov experiment located in the Kamioka 
mine (Japan), at a depth of $\approx 2700~{\rm m.w.e.}$ (from now on SK).   
This detector has a fiducial volume for supernova neutrinos of 
$32~{\rm kton}$ and is equipped with $11146$ $20~{\rm inch}$ PMTs, 
with a photocathodic coverage of $40 \,\%$ of the total surface. 
An external buffer of water, equipped with $\approx 1800$ $8~{\rm inch}$ 
PMTs, is used as a veto. Energy and angular resolutions at $10~{\rm MeV}$ are 
$\approx 16 \,\%$ and $27^{\circ}$; the energy threshold is $\approx 
6~{\rm MeV}$.\footnote{On $12$ November $2001$ a severe accident destroyed 
about half of the Super-Kamiokande PMTs. The plan of the collaboration 
is to refill the whole detector and to equip it with the surviving 
PMTs; this should degrade the quoted resolutions by about a factor 
$\sqrt{2}$ and raise the energy threshold up to $\sim 8~{\rm MeV}$. 
However, the number of events expected from a galactic stellar collapse 
should be reduced by only few percent.} Table \ref{SKevents} shows the 
expected number of events in SK for a type II supernova at the Galactic 
Center.\footnote{Note that the benchmark model \cite{BUR92} predicts 
no events from ${ }^{16}O$ excitation in SK: this is due to the fact 
that the cross sections for these processes were recently 
evaluated in detail \cite{LAN96} and that the average {\nux} energy 
in this model is $\langle E_{\nu_{x}} \rangle \approx 16~{\rm MeV}$. Other 
calculations, based on harder {\nux} spectra ($\langle E_{\nu_{x}} \rangle 
\approx 25~{\rm MeV}$), predict some hundreds of these events in 
SK for a supernova at the Galactic Center \cite{BEA98a}.}
\begin{table}[htb]
\begin{center}
\caption{Expected number of events in SK ($32~{\rm 
kton}$ of fiducial volume) from a supernova at the Galactic Center.}
\label{SKevents}
\vspace{0.2cm}
\begin{tabular}{|c|c|c|} 
\hline
\hline
\raisebox{0pt}[12pt][6pt]{\bf Reaction}         & 
\raisebox{0pt}[12pt][6pt]{\bf Events} &
\raisebox{0pt}[12pt][6pt]{\bf Fraction (\%)} \\
\hline
\raisebox{0pt}[12pt][6pt]{$\bar{\nu}_e~+~p$}    & 
\raisebox{0pt}[12pt][6pt]{$7349$}               & 
\raisebox{0pt}[12pt][6pt]{$95.9$} \\
\raisebox{0pt}[12pt][6pt]{$\nu_e~+~e$}          & 
\raisebox{0pt}[12pt][6pt]{$107$}                & 
\raisebox{0pt}[12pt][6pt]{$1.4$} \\
\raisebox{0pt}[12pt][6pt]{$\bar{\nu}_e~+~e$}    & 
\raisebox{0pt}[12pt][6pt]{$23$}                 & 
\raisebox{0pt}[12pt][6pt]{$0.3$} \\
\raisebox{0pt}[12pt][6pt]{$\nu_x~+~e$}          & 
\raisebox{0pt}[12pt][6pt]{$69$}                 & 
\raisebox{0pt}[12pt][6pt]{$0.9$} \\
\raisebox{0pt}[12pt][6pt]{$\nu_e~+~O$}          & 
\raisebox{0pt}[12pt][6pt]{$50$}                 & 
\raisebox{0pt}[12pt][6pt]{$0.65$} \\
\raisebox{0pt}[12pt][6pt]{$\bar{\nu}_e~+~O$}    & 
\raisebox{0pt}[12pt][6pt]{$63$}                 & 
\raisebox{0pt}[12pt][6pt]{$0.85$} \\
\hline
\raisebox{0pt}[12pt][6pt]{\bf Total on e}       & 
\raisebox{0pt}[12pt][6pt]{$199$}                & 
\raisebox{0pt}[12pt][6pt]{$2.6$}  \\
\raisebox{0pt}[12pt][6pt]{\bf Total on O}       & 
\raisebox{0pt}[12pt][6pt]{$113$}                & 
\raisebox{0pt}[12pt][6pt]{$1.5$}  \\
\raisebox{0pt}[12pt][6pt]{\bf Total on p}       &  
\raisebox{0pt}[12pt][6pt]{$7349$}               & 
\raisebox{0pt}[12pt][6pt]{$95.9$} \\
\hline
\raisebox{0pt}[12pt][6pt]{\bf Total}            & 
\raisebox{0pt}[12pt][6pt]{$7661$}               & 
\raisebox{0pt}[12pt][6pt]{$100$} \\
\hline
\hline
\end{tabular}
\end{center}
\end{table}

\subsection{Heavy water \v{C}erenkov detectors}\label{sec:44}
\begin{figure}[htb]
\begin{center}
\mbox{
	\psfig{file=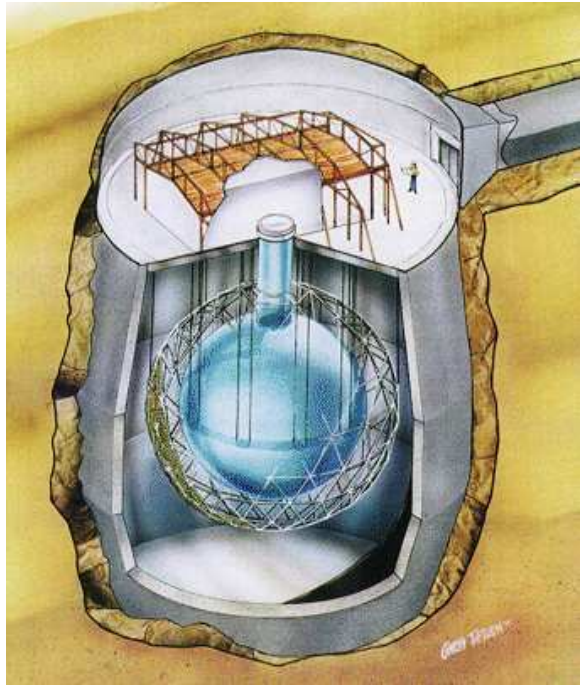,width=3.0in}
     }
\end{center}
\caption{The SNO experiment.}
\label{SNO}
\end{figure}

{\bf SNO} ({\bf S}udbury {\bf N}eutrino {\bf O}bservatory) \cite{SNO00} 
(Fig. \ref{SNO}) is a large heavy water \v{C}erenkov detector, located 
in the Creighton mine (Canada), at a depth of $6010~{\rm m.w.e.}$. The 
experiment is based on $1000~{\rm tonn}$ of $D_{2}O$, surrounded by an 
external shield of $5000~{\rm tonn}$ of light water (the inner 
$1400~{\rm tonn}$ of water can also be used for supernova neutrino 
detection).   
The heavy water is looked by $> 9000$ PMTs and the light water 
by $\approx 2000$ PMTs. The energy and angular resolutions at supernova 
neutrino energies are close to that of SK; the energy 
threshold is $\sim 4~{\rm MeV}$.  

The simultaneous presence of heavy and light water makes SNO a very 
promising experiment for detecting supernova neutrinos, since it has a  
good sensitivity to neutrinos of all flavours. This versatility comes 
mainly from the low-threshold Deuterium break-up reactions (\ref{nued}), 
(\ref{anued}) and (\ref{nudisin}). To have a good efficiency in capturing the 
neutron emitted in (\ref{nudisin}), a $NaCl$ doping will be added 
to the heavy water. The Chlorine has a high cross section for neutron 
capture; its de-excitation (with a time constant of $4~{\rm ms}$) 
releases a $\approx 8.6~{\rm MeV}$ electromagnetic 
cascade. The capture efficiency with the planned $NaCl$ doping is $\epsilon 
\approx 83 \, \%$ \cite{WAL01}. Note that the $\gamma$ background 
is a potentially serious problem for the NC measurement, since what 
a neutrino disintegrates, a $\gamma$ disintegrates too. 
The measured $U$ and $Th$ radioactive contaminations \cite{MDO00},  
at a level of few times $10^{-14}~{\rm g}/{\rm g}$, indicate that 
the NC reactions can be observed, even for solar neutrinos, with 
small systematic uncertainties. 

Table \ref{SNOevents} shows the expected number of events in SNO 
for a supernova at the Galactic Center.\footnote{As already observed 
for SK, the \cite{BUR92} model does not take into account NC reactions 
on Oxygen; more recent papers \cite{BEA98b} predict $\approx 60$ events 
of this type in SNO for a supernova at the Galactic Center.}  
\begin{table}[htb]
\begin{center}
\caption{Expected number of events in SNO from a supernova 
at the Galactic Center.} 
\label{SNOevents}
\vspace{0.2cm}
\begin{tabular}{|c|c|c|}
\hline
\hline
\raisebox{0pt}[12pt][6pt]{\bf Reaction}         & 
\raisebox{0pt}[12pt][6pt]{\bf Events} & 
\raisebox{0pt}[12pt][6pt]{\bf Fraction (\%)} \\
\hline
\hline
\raisebox{0pt}[12pt][6pt]{$\bar{\nu}_e + p$}      &     
\raisebox{0pt}[12pt][6pt]{$446$}                  & 
\raisebox{0pt}[12pt][6pt]{$39.8$} \\
\raisebox{0pt}[12pt][6pt]{$\nu_e + e$}            & 
\raisebox{0pt}[12pt][6pt]{$26$}                   & 
\raisebox{0pt}[12pt][6pt]{$2.3$} \\
\raisebox{0pt}[12pt][6pt]{$\bar{\nu}_e + e$}      & 
\raisebox{0pt}[12pt][6pt]{$8$}                    & 
\raisebox{0pt}[12pt][6pt]{$0.7$} \\
\raisebox{0pt}[12pt][6pt]{$\nu_x + e$}            & 
\raisebox{0pt}[12pt][6pt]{$12$}                    & 
\raisebox{0pt}[12pt][6pt]{$1.1$} \\
\raisebox{0pt}[12pt][6pt]{$\nu_e + O$}            & 
\raisebox{0pt}[12pt][6pt]{$4 \div 5$}             & 
\raisebox{0pt}[12pt][6pt]{$0.36 \div 0.45$} \\
\raisebox{0pt}[12pt][6pt]{$\bar{\nu}_e~+~O$}      & 
\raisebox{0pt}[12pt][6pt]{$5 \div 6$}             & 
\raisebox{0pt}[12pt][6pt]{$0.45 \div 0.54$} \\
\raisebox{0pt}[12pt][6pt]{$\nu_e + d~(CC)$}       & 
\raisebox{0pt}[12pt][6pt]{$113$}                  & 
\raisebox{0pt}[12pt][6pt]{$10.1$} \\
\raisebox{0pt}[12pt][6pt]{$\bar{\nu}_e + d~(CC)$} & 
\raisebox{0pt}[12pt][6pt]{$201$}                   & 
\raisebox{0pt}[12pt][6pt]{$17.9$} \\
\raisebox{0pt}[12pt][6pt]{$\nu_e + d~(NC)$}       & 
\raisebox{0pt}[12pt][6pt]{$43$}                   & 
\raisebox{0pt}[12pt][6pt]{$3.8$} \\
\raisebox{0pt}[12pt][6pt]{$\bar{\nu}_e + d~(NC)$} & 
\raisebox{0pt}[12pt][6pt]{$44$}                   & 
\raisebox{0pt}[12pt][6pt]{$3.9$} \\
\raisebox{0pt}[12pt][6pt]{$\nu_{x} + d~(NC)$}     & 
\raisebox{0pt}[12pt][6pt]{$224$}                  & 
\raisebox{0pt}[12pt][6pt]{$20.0$} \\
\hline
\raisebox{0pt}[12pt][6pt]{\bf NC on D}            & 
\raisebox{0pt}[12pt][6pt]{$311$}                  & 
\raisebox{0pt}[12pt][6pt]{$27.7$} \\
\raisebox{0pt}[12pt][6pt]{\bf CC on D}            & 
\raisebox{0pt}[12pt][6pt]{$314$}                  & 
\raisebox{0pt}[12pt][6pt]{$28.0$} \\
\raisebox{0pt}[12pt][6pt]{\bf CC on O}            & 
\raisebox{0pt}[12pt][6pt]{$10$}                   & 
\raisebox{0pt}[12pt][6pt]{$0.9$} \\
\raisebox{0pt}[12pt][6pt]{\bf Total on e}         & 
\raisebox{0pt}[12pt][6pt]{$46$}                   & 
\raisebox{0pt}[12pt][6pt]{$4.1$} \\
\raisebox{0pt}[12pt][6pt]{\bf Total on p}         & 
\raisebox{0pt}[12pt][6pt]{$446$}                  & 
\raisebox{0pt}[12pt][6pt]{$39.8$} \\
\hline
\raisebox{0pt}[12pt][6pt]{\bf Total on $D_2O$}    & 
\raisebox{0pt}[12pt][6pt]{$645$}                  & 
\raisebox{0pt}[12pt][6pt]{$57.5$} \\
\raisebox{0pt}[12pt][6pt]{\bf Total on $H_2O$}    & 
\raisebox{0pt}[12pt][6pt]{$476$}                  & 
\raisebox{0pt}[12pt][6pt]{$42.5$} \\
\hline
\raisebox{0pt}[12pt][6pt]{\bf Total}             & 
\raisebox{0pt}[12pt][6pt]{$1121$}                & 
\raisebox{0pt}[12pt][6pt]{$100$} \\
\hline
\hline
\end{tabular} 
\end{center}
\end{table}
\subsection{Other detectors}\label{sec:45}
\subsubsection{ICARUS}
{\bf ICARUS} ({\bf I}maging of {\bf C}osmic {\bf A}nd {\bf R}are 
{\bf U}nderground {\bf S}ignals) is a 
modular Liquid Argon projection chamber; the first module, with a 
sensitive mass of $600~{\rm tonn}$, will be installed soon in the 
LNGS (Hall \lq\lq C\rq\rq).
The Liquid Argon is mainly sensitive to {\nue}, via the 
(\ref{ArLi}) process; $\sim 40$ {\nue} events could be detected in one 
ICARUS module  
for a supernova at the Galactic Center, with reasonable chances to 
have a good signature of the infall-neutronization burst. Some other 
($\sim 10$) events are expected by ES reactions induced by neutrinos of 
all flavours \cite{ICA01}.
\subsubsection{High Energy Neutrino Telescopes (HENTs)}
{\bf HENT}s ({\bf AMANDA} and {\bf Baikal} (already on-line), {\bf NESTOR}, 
{\bf ANTARES} and {\bf NEMO} (under development)) are arrays of hundreds 
of Optical Modules ({\it OM}), deployed in long strings ($\sim {\rm Km}$) 
in deep sea or antarctic ice. 
These detectors look at energetic cosmic ray muons and have an energy 
threshold in the GeV range; however, they have some sensitivity to a galactic 
supernova explosion because of a collective effect, the excess of single 
counting rates produced in all OMs by a stream of thousands of low energy 
positrons concentrated in a short time \cite{HAL94}. A supernova trigger 
based on this idea is active in {\bf AMANDA} ({\bf A}ntartic {\bf M}uon 
{\bf A}nd {\bf N}eutrino {\bf D}etector {\bf A}rray) \cite{NEU01}; 
the experimental measurements of environmental background and PMT noise 
and the Monte Carlo calculations of the expected signal \cite{HAL96} showed 
that AMANDA can detect galactic supernova explosions with a statistical 
significance $\ge 6~\sigma$ above the average counting rate.   
\subsubsection{Radiochemical detectors}
The radiochemical detectors, like {\bf GNO}, {\bf SAGE} and {\bf Homestake}, 
are time-integrated detectors, sensitive to solar {\nue}'s via CC reactions. 
The active material is periodically ($T \sim 1~{\rm month}$) extracted to 
look for isotopes produced by the CC reactions. Their value as supernova 
detectors is clearly rather limited, because no information can be 
extracted about timing, energy and direction of neutrinos. However, in 
case of a nearby supernova, some {\nue} events should be recorded by 
these detectors and a prompt extraction could allow to determine 
whether a statistically significant increase of the counting rate 
was observed in the relevant period.   
\subsection{Ideas for \lq\lq far\rq\rq~future projects}\label{sec:46}
The experimental success of detectors like SK or 
AMANDA stimulated some ideas to extend the technologies employed 
in these experiments to much larger scales. The goals of such 
projects is to gain order-of-magnitudes with respect to the present 
sensitivities on proton decay search, detection of solar, atmospheric 
and high energy neutrinos etc.; moreover, the long baseline and neutrino 
factory projects demand very large and refined neutrino detectors.    

Here I look at some of these ideas from the point of view of the 
supernova neutrino detection.  
\subsubsection{UNO}
{\bf UNO} ({\bf U}ltra underground {\bf N}ucleon decay and neutrino 
{\bf O}bservatory) is a project for a large scale Water \v{C}erenkov 
detector. The idea is to reach a total mass of $650~{\rm kton}$ of water, 
organized in three compartments, $60 \times 60 \times 60~{\rm m}^3$ each, 
arranged in a linear structure. This configuration seems the most promising 
(in respect, e.g., with a cubic shape) in terms of fiducial 
volume, duty cycle, mechanical stability etc. The fiducial volume should 
be $445~{\rm kton}$ ($14$ times that of SK). The total number 
of PMTs is $\sim 60000$ (with a photocatodic coverage of $40 \, \%$ in the 
central section and $10 \, \%$ in the side ones) and the estimated cost 
is $500~{\rm M\$}$. A galactic stellar collapse should produce 
$\sim~10^{5}$ $e^{+}$ events in UNO, but also a supernova in the Andromeda 
galaxy (the closest to the Milky Way, at a distance of $700~{\rm Kpc}$) 
should be observable in this detector, with an expected signal of few 
tenths of events. The sensitivity to extra-galactic supernov\ae~is 
an important quality factor, since the rate of stellar collapses in 
the Local Group is expected to be several times higher than that 
in our galaxy; this gives a much higher chance of success to an 
experiment capable to look beyond the Milky Way \cite{JUN00}.  
\subsubsection{SNBO/OMNIS, LAND}
{\bf SNBO/OMNIS}~({\bf S}upernova~{\bf N}eutrino~{\bf B}urst~{\bf 
O}bser\-va\-tory; the {\bf O}bservatory for {\bf M}ultiflavour 
{\bf N}eutrinos from {\bf S}upernov\ae) (\cite{CLI90}, \cite{SMI97}) 
and {\bf LAND} ({\bf L}ead {\bf A}stronomical {\bf N}eutrino {\bf D}etector) 
\cite{HAR96} are projects of large mass ($\sim 10^{4}~{\rm tonn}$) 
detectors, made of low-cost, high-$Z$ materials ($NaCl$, $Fe$ and/or 
$Pb$), whose main goal is the observation of NC events from a supernova 
neutrino burst. The neutrons emitted in knocking-off NC processes should 
be observed by long ${ }^{6}Li$ or ${ }^{10}B$ detectors or by 
$Gd$-doped liquid scintillation counters, interspaced within the target 
materials. These detectors should record, in case of a galactic supernova, 
$\sim 1000$ NC events, mainly induced by {\nux}'s. The OMNIS collaboration 
is also considering a different detector design, based on $2~{\rm kton}$ 
of Lead perchlorate, with a high sensitivity to {\nue} too. Experiments 
based on neutron spallation reactions require a neutron poor 
environment; recent measurements seem to indicate 
the Carlsbad (New Mexico) site as a promising one \cite{CLI99}.   

\subsubsection{LANNDD}
{\bf LANNDD} ({\bf L}iquid {\bf A}rgon {\bf N}eutrino and {\bf N}ucleon 
{\bf D}ecay {\bf D}etector) is a proposal of a $70~{\rm kton}$ magnetized 
Liquid Argon tracking detector. In case of a galactic supernova this detector 
should observe $\sim 3000$ {\nue} CC reactions (\ref{ArLi}) \cite{CLI01}.
\subsubsection{IceCube}
{\bf IceCube} is a project of a $1~{\rm Km}^3$ volume neutrino telescope, 
to be located at the South Pole, designed for the detection of 
extremely high energy neutrinos ($\gsim~10^{20}~{\rm eV}$) of astrophysical 
origin. This experiment should 
be equipped with a supernova trigger of the AMANDA type, with improved 
sensitivity due to the larger ($4800$) number of OMs. IceCube should be 
able to identify the leading front of the supernova neutrino signal with 
a $\lsim \; 3~{\rm ms}$ absolute timing accuracy, instead of the $15~{\rm ms}$ 
AMANDA accuracy. This number looks promising from the point of view of 
supernova direction reconstruction by the triangulation method (see 
section \ref{sec:53}) \cite{GOL01}.
\section{What can we learn ?}\label{sec:5}
A complete review of all that might be learned from a supernova explosion 
would require the time and space of a very long report; then I shall limit 
myself to few points. 
\subsection{Determination of supernova and neutrino parameters}\label{sec:51}
The large neutrino burst that is expected from a galactic supernova 
should allow the extraction of some parameters of the supernova source 
and of the neutrino signal. Here I do not discuss how these determinations 
are affected by possible neutrino oscillations; this point will be examined 
in section \ref{sec:54}. 

Present detectors (expecially SK) should record thousands of (\ref{anuep}) 
events; this will allow a high accuracy measurement of the 
{\anue} spectrum and time profile. A simple fit with trial spectra to 
these data should provide the {\anue} temperature $T_{\bar{\nu}_{e}}$ 
and chemical potential $\mu_{\bar{\nu}_{e}}$ with a $1 \, \%$ accuracy. 
A lower accuracy ($\sim 10 \, \%$) measurement should be possible also 
for {\nue}, via CC (p.e. on Deuterium) and ES reactions.  
The total emitted energy $E_{B}$ could also be measured, since it is 
related to the total number of events in both flavours and to the 
supernova distance $D$; the two independent measurements of $E_{B}/D^{2}$ 
(the source strength at the detector) could 
be compared to check whether the energy equipartition hypothesis holds. 
 
The extraction of $T_{\nu_{x}}$ is more complicated, since the NC reactions 
do not preserve any information on the neutrino energy. However, if one 
assumes the energy equipartition, an estimation of $T_{\nu_{x}}$ will 
be obtained by comparing the observed number of events with that predicted 
from a supernova of the measured $E_{B}/D^{2}$. Note that any separation 
between $\nu_{\mu}$ or $\nu_{\tau}$ is forbidden by the nature of the 
detection process (the NC reactions are insensitive to the flavour). The 
rough estimation of $T_{\nu_{x}}$ will also allow to test the 
temperature hierarchy: 
$T_{\nu_{x}} > T_{\bar{\nu}_{e}} > T_{\nu_{e}}$. 

Note that the energy spectrum and the total number of events are not 
affected by a (possible) finite neutrino mass; then, the 
temperature and source strength determinations discussed above are 
valid for massless and for massive neutrinos too. Only in case of a 
\lq\lq large\rq\rq~($\sim$ some ${\rm KeV}$) neutrino mass a potential 
problem is present: since the neutrino signal would be largely broadened, 
some events could become indistinguishable from the time-independent 
background and would be lost. This would produce an underestimation 
of the total number of events and then a normalization problem.
\subsection{Fast supernova observation: the SuperNova Early Warning System 
(SNEWS)}\label{sec:52}
The astrophysical models predict, as confirmed by the SN$1987$A observation, 
that the neutrino signal preceeds the supernova optical flare by some hours, 
the time needed to the shock wave to propagate through the collapsed 
matter and to the optical light to reach observable magnitudes. 
Astronomers are very interested in observing the first light 
from a supernova (not easy to do in case of an extragalactic supernova); 
this light carries information on the supernova progenitor and its 
immediate environment. So, many experiments developed systems for prompt 
recognitions of neutrino bursts from supernov\ae. 

\begin{figure*}[htb]
\begin{center}
\mbox{
\psfig{figure=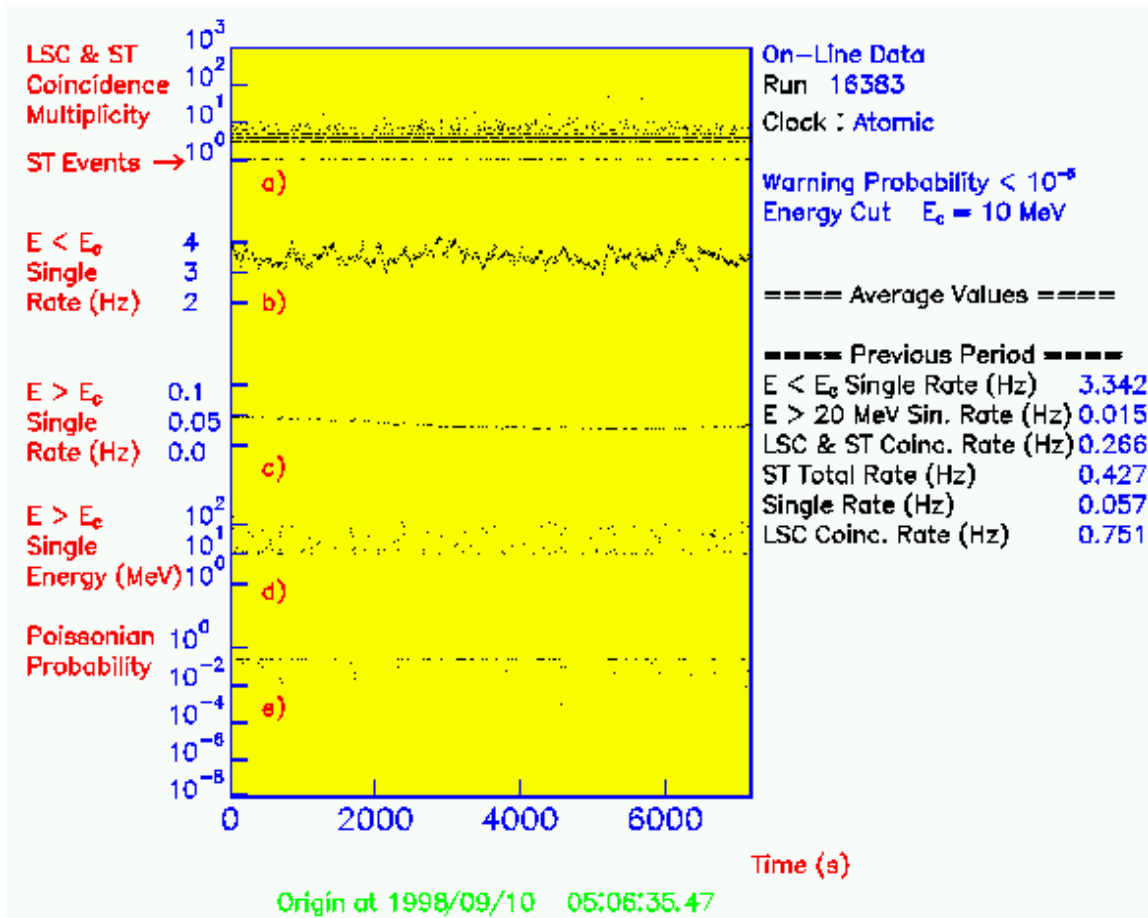,width=6.0in}
}
\caption{Display of the supernova monitor program of the MACRO 
experiment (from [MAC98a]).}
\label{macmon}
\end{center}
\end{figure*}

All these systems are based on the pulsed character of the supernova 
neutrino signal: given the normal trigger rate of an experiment (due to 
the environmental and cosmic background), a neutrino burst should 
produce a fast increase of this trigger rate, well above any poissonian 
statistical fluctuation. If such a signal is observed, a second level 
analysis is performed to recognize whether it matches the expected 
characteristics for a genuine neutrino burst.  
Fig. \ref{macmon} shows the display of the supernova monitor program which 
was running in the MACRO experiment. 
A possible burst was signalled by a very low probability of poissonian 
fluctuations of the background rate (the last scale from above). 
In the other scales the behaviour of the apparatus was monitored, 
looking at physical quantities as the energy of the events 
observed in the scintillation counters.
Similar systems are operating in SK, LVD, SNO and AMANDA. 

\begin{figure}[htb]
\begin{center}
\mbox{
\psfig{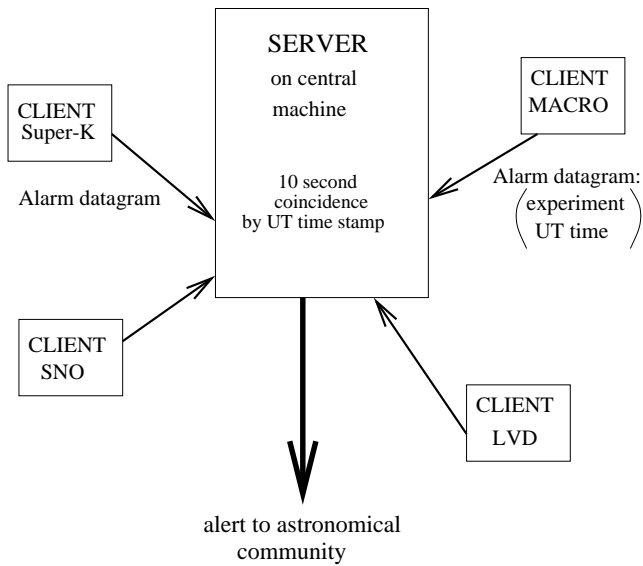}
}
\caption{The SNEWS setup (from [SCH99]).}
\label{snewsset}
\end{center}
\end{figure}

All these systems are members of a coordinate network of supernova 
observatories (the {\bf SNEWS}, {\bf S}upernova {\bf N}etwork 
{\bf E}arly {\bf W}arning {\bf S}ystem), whose goal is to provide 
a fast alert to the astronomical observatories around the world. 
This system is based on a blind computer, which gets separate alerts 
from any participating experiment and looks for possible time 
coincidences, in a $10~{\rm s}$ window, between such alerts. 
This procedure eliminates the human interventions needed to check 
the supernova-like nature of the burst (which produces a significant 
loss of time for observation) and ensures a very high level of confidence, 
since an accidental coincidence of fake supernova bursts (due to poissonian 
fluctuations or detector pathologies) between different experiments 
at distances of thousands of Km is extremely unlike \cite{SCH99}. 
The SNEWS setup is shown in Fig. \ref{snewsset}.
The estimated response time of SNEWS is few tenths of minutes.  
\subsection{Identification of supernova direction}\label{sec:53}
The prompt recognition of a supernova burst is not 
the unique information that supernova neutrino detectors should provide 
to astronomical observatories, since it is not easy to find something in 
the sky without any guess on where to look. The pointing back to 
the supernova is possible with two different techniques: the angular 
distributions of some neutrino detection reactions and the triangulation, 
which takes advantage from the presence of SNEWS.
\subsubsection{Angular distributions.}
The most promising reaction for this purpose is the ES (\ref{esnue}) process. 
The extensive discussion \cite{BEA99a} shows that, with some hundreds 
of ES events, SK should be able to identify the supernova direction 
with a $\delta \theta \approx 5^{\circ}$ accuracy for a collapse 
at the Galactic Center. The accuracy is determined by the intrinsic 
angle between the recoil electron and the incident neutrino, by the width 
of the \v{C}erenkov cone, by the multiple scattering of the electrons in 
water and by the uncertainties introduced by the statistical 
separation of the ES events from the dominant (\ref{anuep}) signal. The 
result is given under very pessimistic assumptions on the separation 
capabilities; using more refined techniques this accuracy can be improved 
down to $3 \div 4^{\circ}$. A lower accuracy (due to the limited 
statistics) localization can be obtained in Sudbury: 
$\delta \theta \approx 20^{\circ}$. These results are largely independent 
from the details of the supernova model and particularly of the 
time evolution of the neutrino luminosity. 
 
The angular distributions of the (\ref{anuep}) process and of the CC reactions 
on Oxygen and Deuterium can provide information at a lower level, but they 
can be used as a check of the higher accuracy results based on ES 
reactions to confirme, at least, the sky emisphere where to look.   

As observed before, scintillation detectors are sensitive to $\gamma_{2.2}$, 
while water \v{C}erenkov experiments are not. This opens to the scintillation 
detectors an other opportunity to provide information on the supernova 
direction. The positron is detected nearly at the point of creation, while 
the neutron is boosted forward. The neutron thermalization (dominated by 
elastic scatterings on protons) preserves 
the initial direction in the first collisions and then becomes isotropic.
The $\gamma_{2.2}$ emission is an isotropic process and the $\gamma_{2.2}$ 
position is reconstructed with uncertainties of few tenths of {\rm cm}; 
these degrading effects reduce the significance of the initially 
forward motion, but do not completely destroy the memory of it.  
As a result, the reconstructed point of neutron capture is, 
on the average, displaced by a small amount with respect to the positron 
position. The average displacement is few {\rm cm}, a difference which 
can be measured because of the good localization capabilities of the 
liquid scintillation detectors. The CHOOZ experiment \cite{CHO00a} measured 
a $1.5~{\rm cm}$ average displacement for low energy ($E \; \lsim \; 
8~{\rm MeV}$) {\anue}'s from a nuclear reactor and the reactor direction was 
reconstructed, using this technique, with a $18^{\circ}$ 
accuracy. At supernova neutrino energies the higher kinetic energy of the 
neutron and the lower neutron-proton cross section should enhance the effect. 
The CHOOZ collaboration estimated that a liquid scintillation detector 
with a mass like that of SK would reach a $9^{\circ}$ pointing 
accuracy, a result worse within a factor $2$ than that obtainable 
using ES events.   
\subsubsection{Triangulation}  
The triangulation technique uses the difference in arrival times of 
neutrino signals in various detectors around the world to determine the 
supernova direction. If $\Delta t$ is the difference in arrival time 
of the signals and $d$ is the physical distance between two detectors 
(in time units, $d \approx 40~{\rm ms}$ for detectors located on the 
opposite sides of the earth), the angle $\theta$ between the axis 
connecting the experiments and the supernova direction is given by:
\begin{equation}
\cos \theta = \frac{\Delta t}{d}
\label{triang}
\end{equation}
The uncertainty of this technique is dominated by the uncertainty in 
measuring $\Delta t$, which is related to the capabilities of the two 
experiments of timing the leading edge of the neutrino signal.  
The accuracy obtainable for a collapse at the Galactic Center using 
a triangulation between SK and SNO ($d \approx 30~{\rm ms}$) 
is evaluated in \cite{BEA99b} under reasonable and under extreme 
assumptions on the event time profile (note that a high statistics 
detector like SK could measure this time profile in detail). 
The lower statistics of the SNO signal determines a maximum accuracy 
$\delta \left( \Delta t \right) \approx 15~{\rm ms}$, so that the 
best obtainable accuracy on the cosine of the angle is 
$\delta \left( \cos \theta \right) \approx 0.5$ ($0.25$ only in the 
unrealistic hypothesis of a zero rise-time signal).
If one considers triangulations between SK, SNO and one of the other 
present detectors (e.g. LVD or AMANDA) there are no substantial 
advantages, since the uncertainties in timing the start of the  
neutrino signal for these experiments are at the level of 
that of SNO. An important improvement is expected from the application 
of the triangulation technique to (possible) future detectors, with 
higher statistics and sensitivity. For example, a three detectors 
triangulation involving SK, SNO and IceCube should locate the 
supernova direction with an uncertainty varying from 
$5^{\circ}$ to $20^{\circ}$ \cite{NEU01}.  
\subsection{Non-standard neutrino physics with supernov\ae}\label{sec:54}
A galactic supernova would provide a unique opportunity for searching  
for non standard model properties of neutrinos. Here I discuss how 
a supernova neutrino signal can give important insights on neutrino 
masses and oscillations.
\subsubsection{$\nu_{e}$ mass}
The present {\anue} mass limit, obtained by Tritium $\beta$-decay experiments, 
is $\approx 3~{\rm eV}$ (\cite{WEI99}, \cite{LOB99}). In case of a 
galactic supernova at a distance $D$, a massive neutrino of energy 
$E$ should arrive on the earth with a delay:
\begin{equation}
\Delta t \left( E, m \right) \, \left( s \right) = 0.515 \, \left(\frac
      {m \left( {\rm eV} \right)} {E \left( {\rm MeV} \right)} 
      \right)^2 \, D \left( 10~{\rm Kpc} \right)
\label{massdelay}
\end{equation}
with respect to a massless neutrino emitted by the supernova at the same 
time. The intrinsic duration of the cooling phase ($\approx 10~{\rm s}$) 
tends to mask the delay induced by a finite neutrino mass and determines 
the minimum {\nue} mass whose effects can be explored by an experiment. 
In case of SN$1987$A, model dependent and independent limits were obtained 
in the range $11 \div 23~{\rm eV}$ (see \cite{BAH89} and references 
therein). An other more sensitive approach was recently suggested 
\cite{TOT98b} which, taking advantage from the high statistics expected 
in present supernova detectors (expecially SK), avoids this problem 
using the events ($\sim 300$ in SK) observed in a short time window, 
the first $\sim 50 \div 100~{\rm ms}$ after the explosion. The basic 
idea is that, in case of massless neutrinos, the neutrino arrival times 
$t_i$ and energies $E_i$ are essentially uncorrelated, since in a so short 
time window the energy spectrum does not change significantly. The correlation 
introduced by a finite neutrino mass (equation (\ref{massdelay})) is removed 
if one considers the modified time sequence ${t_i}^{\prime} = t_i - 
\Delta t \left( E_i, m \right)$. Using groups of trial neutrino masses 
$m_{\bar{\nu}_{e}}$ and temperatures $T_{\bar{\nu}_{e}}$, one can evaluate 
the degree of uncorrelation between ${t_i}^{\prime}$ and $E_i$ by statistical 
methods; the maximum uncorrelation is obtained for the correct values of 
$m_{\bar{\nu}_{e}}$ and $T_{\bar{\nu}_{e}}$. The expected sensitivity of this 
method is down to $3~{\rm eV}$, at the same level of the present terrestrial 
limit, and results largely model independent.       
\subsubsection{$\nu_{\mu}$ and $\nu_{\tau}$ masses}
The terrestrial limits of $\nu_{\mu}$ and $\nu_{\tau}$ masses (properly 
speaking, of the predominant mass eigenstates of them) are $170~{\rm KeV}$ 
\cite{ASS96} and $18~{\rm MeV}$ \cite{ALE98}, hard to significantly 
improve with usual techniques; however, cosmological bounds suggest 
\cite{RAF96} that these masses could not exceed tenths of {\rm eV}. 
A galactic supernova should produce hundreds of {\nux} events in the 
present detectors. Several ideas were therefore proposed in the past 
(see e.g. \cite{ACK90}, \cite{RYA92}) to set stringent {\nux} mass 
limits using a supernova neutrino burst. 

Such techniques are based on the statistical separation 
of the neutral current signal from the dominant one (\ref{anuep}) 
in detectors (mainly liquid scintillators) sensitive to both. The 
feasibility of such a separation determines the minimum exploitable 
{\nux} mass; sensitivities down to $100 \div 150~{\rm eV}$ masses  
were obtained. The maximum exploitable {\nux} mass $m_{max}$ is set 
by the level of the experimental background, since for very high masses 
the {\nux} burst becomes so broad that the average time distance between 
the events is comparable with the inverse of the normal trigger rate. 
For present detectors $m_{max} \sim 1 \div 10~{\rm KeV}$. 

Other techniques (\cite{KRA92}, \cite{FIO97}), developed for water 
\v{C}erenkov detectors, use the high directionality of the ES reactions 
to form two samples of event, one made almost only by {\nue} and {\anue}, 
the other containing also ${\nu_{x}}$-induced events. The first sample is 
used to perform a statistical subtraction of the irriducible {\nue}, 
{\anue} background from the second one; the time profile obtained 
after this subtraction is affected by a finite neutrino mass. The minimum 
detectable mass is $\sim 50~{\rm eV}$ in \cite{KRA92} and 
$> 100~{\rm eV}$ in \cite{FIO97}, a difference ascribed to the very 
quickly decaying neutrino luminosity of the model used in the former paper.   

A recently proposed technique \cite{BEA98b} enhances the sensitivity 
to small neutrino masses. 

\begin{figure}[htb]
\begin{center}
\mbox{
	\psfig{file=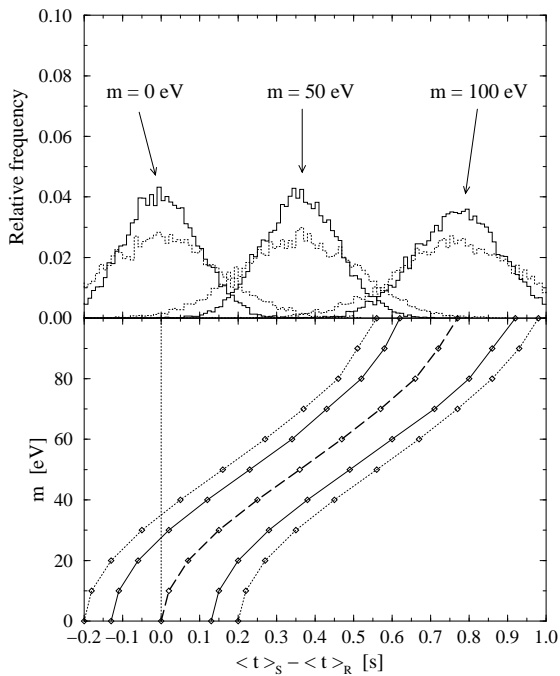,height=4.0in}
	}
\end{center}
\vspace{-1cm}
\caption{Mass sensitivity analysis for SNO. Upper part: distribution of 
the time difference $\langle t_S \rangle - \langle t_R \rangle$ for three 
representative cases. The solid line is for reference time profile extracted 
from SK data, the dotted line for reference extracted from SNO data. Lower 
part: the range of masses corresponding to a given measured 
$\langle t_S \rangle - \langle t_R \rangle$. The dashed line is the 
corresponding central value; the continuous lines are $10 \, \%$ and 
$90 \, \%$ C.L. limits obtained using the SK $R \left( t 
\right)$ and the dotted lines are the same using the SNO $R \left( t 
\right)$ (from [BEA98b]).}
\label{SNOsens}
\end{figure}

Let us assume that {\nue} (and then {\anue}) are massless, while 
{\nutau} and {\numu} (at least one) are massive. The equation 
(\ref{massdelay}) can not be immediately applied to extract $m$ because 
of the neutrino signal duration and since the {\nux} energy is not 
measured (except that in ES reactions). However, consider two 
samples of events, one ($R \left( t \right)$, the \lq\lq Reference\rq\rq) 
formed by massless neutrinos only and one formed mainly by massive 
neutrinos, with some contaminations from massless neutrinos 
($S \left( t \right)$, the \lq\lq Signal\rq\rq). The first sample could be 
extracted by SK (or LVD) {\anue} data 
or by the heavy water portion of the SNO signal and 
the second by ${ }^{16}O$ NC reactions in SK, or by NC reactions on 
Deuterium in SNO etc. The residual contamination (estimated $\sim 20 
\,\%$) comes from the dominant {\anue} signal in the energy interval 
$5 \div 10~{\rm MeV}$ for SK and from CC reactions on Deuterium for SNO; 
in both cases, an event-by-event separation is not possible. For each 
of these samples, one can define an average arrival time ($\langle t_R 
\rangle$ and $\langle t_S \rangle$) and compare the two values. The 
signature of a finite {\nux} mass is given by a statistically 
significant difference $\langle t_S \rangle - \langle t_R \rangle > 0$. 
Fig. \ref{SNOsens} shows the results of $10^{4}$ Monte Carlo 
simulations (each simulation is a supernova) of the difference 
$\langle t_S \rangle - \langle t_R \rangle$ \cite{BEA98b} in the case 
of SNO; the supernova distance is taken as $10~{\rm Kpc}$ and the 
samples are generated taking into account the number 
of events expected in the various detectors.
The upper part of the plot shows the distribution of the difference 
$\langle t_S \rangle - \langle t_R \rangle$ for three representative 
cases ($m = 0, \, 50, \, 100~{\rm eV}$); the distributions 
are narrower if the reference signal $R \left( t \right)$ is taken from 
the larger statistics sample provided by SK and wider if, instead, 
$R \left( t \right)$ is extracted from the SNO data. The lower part is 
a sort of \lq\lq reverse\rq\rq~plot: given a measured 
$\langle t_S \rangle - \langle t_R \rangle$ difference, 
the thick dashed line defines the corresponding central {\nux} mass 
value, while the continuous lines define the $10 \, \%$ and $90 \, \%$ C.L. 
limits obtained using SK data for $R \left( t \right)$; the 
dotted lines are the same if one uses the SNO data for $R \left( t \right)$. 
This figure shows that the SNO sensitivity is expected to reach 
$\sim 30~{\rm eV}$, a six orders-of-magnitude improvement for the 
$\nu_{\tau}$ mass limit. 

Similar analyses performed for other detectors predict sensitivities down 
to $50~{\rm eV}$ for SK, $55~{\rm eV}$ for Kamland and $75~{\rm eV}$ for 
Borexino (\cite{BEA99b}, \cite{CAD00}); all these results are obtained 
assuming that only $\nu_{\tau}$ is massive. However, the claimed evidence 
of atmospheric neutrino oscillations seems to suggest an almost total mixing 
between $\nu_{\mu}$ and $\nu_{\tau}$ \cite{SKA98b}; if this is the case, 
the quoted sensitivities improve by a factor $\sqrt{2}$. 

The results discussed here are largely independent from the supernova 
distance (within our galaxy) since the smaller delay is compensated 
by the increased statistics. Note that this technique uses the large 
{\anue} signal as an internal clock; this makes the results less 
dependent on the model, since it does not require theoretical 
assumptions on the time pattern of the neutrino signal, except that 
the cooling phase duration is $\sim 10~{\rm s}$, as observed for SN$1987$A. 
A much longer (hundreds of seconds or more) cooling phase would make 
the analysis more complicated and less sensitive, since the experimental 
background might not be neglected, as implicitly assumed. 

The potentialities of the proposed NC sensitive detectors based on 
heavy nuclei (SNBO/OMNIS and LAND) for the {\nux} mass measurement 
were also discussed (\cite{CLI94}, \cite{SMI97}). The {\nux} mass 
value should be inferred from the distorsions of the observed time 
profile from that predicted for a massless particle (the neutrino 
massless time profile is measurable with high accuracy by SK); 
sensitivities down to few tenths of {\rm eV} are claimed.

\vspace{0.1cm}
Even better sensitivities could be obtained in the particular case 
of a supernova which rapidly degenerates into a black hole \cite{BEA00}. 
In this case the neutrino signal would terminate abruptly at the 
black hole formation time $t_{BH}$, with an expected transition of 
$\lsim~0.5~{\rm ms}$ duration. In case of a finite neutrino mass, the 
drop in luminosity would be steeper for higher energy neutrinos and 
smoother for lower energy neutrinos; so, a first sample of energetic 
neutrinos could be used to extract $t_{BH}$ with an accuracy $\lsim~1~{\rm 
ms}$ and a second sample of less energetic neutrinos could be used to 
unfold the neutrino mass value from the tail of the neutrino luminosity 
at $t > t_{BH}$. The expected sensitivity of this technique for the  
present detectors is down to $1.8~{\rm eV}$ for {\nue} and down to 
$6~{\rm eV}$ for {\nux}.    
\subsubsection{Neutrino Oscillations}
The supernova spectra discussed in section \ref{sec:2} are based on 
the hypothesis of massless neutrinos, which do not experience flavour 
mixing. However, the long-standing solar neutrino problem (for 
recent reviews of the problem and of the proposed solutions see e.g. 
\cite{BAH01}, \cite{FOG01}) and the positive indications coming from 
atmospheric neutrino experiments (\cite{SKA98b}, \cite{MAC98b}, 
\cite{SOU97}) are strong hints in favour of neutrino oscillations. An 
uncorfirmed indication comes also from the LSND \cite{LSN98} experiment.
Such effects could manifest during the collapse (expecially if the 
MSW (\cite{MIK86}, \cite{WOL78}) mechanism works) or during the 
neutrino travel from the collapsed star to the earth. Note also that 
a supernova neutrino detection would offer the opportunity to study vacuum 
oscillations over an unprecedented baseline. 

The effects of the neutrino oscillations on the supernova neutrino 
spectra and their signatures in terrestrial detectors were discussed 
by many authors (see e.g. \cite{FUL87}, \cite{BUR92}, \cite{RAF93}, 
\cite{BUR93}, \cite{QIA94}, \cite{FUL99}, \cite{CHO00b}), using a 
large variety of mixing parameters (mass squared differences and 
mixing angles). The neutrino oscillations, for instance, were advocated 
as a strengthening effect for the shock wave neutrino heating \cite{FUL92} 
and the request of an efficient $r$-process nucleosynthesis in the 
supernova core was used as an argument to set bounds on neutrino mixing 
parameters \cite{QIA93}. 

\vspace{0.1cm}
The combination of the results of all neutrino oscillation experiments 
restricts (at $99 \,\%$ C.L.) the possible solutions of the solar 
neutrino puzzle to four regions of the $\left( \Delta m^2, \sin^{2} 2 
\theta \right)$ plane, the vacuum ({\bf VO}) oscillation and the three  
({\bf LMA}, {\bf SMA} and {\bf LOW}) resonant MSW conversions (the 
first one is presently the favoured). Each of them is characterized 
by a different couple of parameters $\left( \Delta m^{2}_{\odot}, 
\sin^{2} 2 \theta_{\odot} \right)$: 
\begin{itemize}
\vspace{-0.1cm}
\item[-] $\left( \left( 4 \div 10 \right) \cdot 10^{-6}~{\rm eV}^2, 
\left(2 \div 10 \right) \cdot 10^{-3} \right)$ for SMA;
\vspace{-0.2cm}
\item[-] $\left( \left( 1 \div 10 \right) \cdot 10^{-5}~{\rm eV}^2, 
\left(0.7 \div 0.95 \right) \right)$ for LMA;
\vspace{-0.2cm}
\item[-] $\left( \left( 0.5 \div 2 \right) \cdot 10^{-7}~{\rm eV}^2, 
\left(0.9 \div 1.0 \right) \right)$ for LOW;
\vspace{-0.2cm}
\item[-] $\left( \left( 6 \div 60 \right) \cdot 10^{-11}~{\rm eV}^2, 
\left(0.8 \div 1.0 \right) \right)$ for VO. 
\end{itemize}
The electron flavour is distributed in the mass eigenstates 
$\nu_{1}$ and $\nu_{2}$, with admixtures given by $U_{e1} \approx \cos 
\theta_{\odot}$, $U_{e2} \approx \sin \theta_{\odot}$. The admixture $U_{e3}$ 
of the electron with the third neutrino mass state is unknown, but is 
strongly bounded by the CHOOZ result \cite{CHO99}: 
$\left| U_{e3} \right|^{2}~\lsim~0.02$. The 
atmospheric neutrino anomaly favours the hypothesis of oscillation between 
two active neutrino states (\lq\lq 2\rq\rq~and \lq\lq 3\rq\rq), usually 
identified with {\numu} and {\nutau}; the corresponding pair of parameters is 
$\left( \Delta m^{2}_{atm}, \sin^{2} 2 \theta_{\mu,\tau} \right)$. The almost 
complete degeneracy between the states is expressed by the condition: 
$\sin^{2} 2 \theta_{\mu,\tau} > 0.88$ and the favoured $\Delta m^{2}_{atm}$ 
value lies in the range 
$\left( 1.5 \div 4 \right) \cdot 10^{-3}~{\rm eV}^{2}$. 

Having these parameters in mind, some general results can be inferred 
\cite{DIG00}:
\begin{itemize}
\vspace{-0.1cm}
\item in the supernova, where the density reaches 
$10^{14}~{\rm g}~{\rm cm}^{-3}$, two different MSW resonances occur. 
Independently of the solar neutrino problem solution, these resonances 
take place in the mantel, without affecting the explosion dynamics and 
expecially the shock neutrino heating. This happens because during 
the $10~{\rm s}$ of the diffusion process the shock wave reaches 
regions where the density is $\gsim~10^{6}~{\rm g}~{\rm cm}^{-3}$, while 
the maximum density for a MSW resonance is 
$\lsim~10^{4}~{\rm g}~{\rm cm}^{-3}$;
\vspace{-0.2cm}
\item the nucleosynthesis is also unaffected, since $\Delta m^{2}_{\odot}$ 
(which determines the {\nue} mixing) is much lower 
than the nucleosynthesis upper bound;
\vspace{-0.2cm}
\item since the MSW effect is sensitive to the sign of $\Delta m^{2}$, 
only {\nue} (in case of normal mass hierarchy) or {\anue} (in case of 
inverted mass hierarchy) can experience a resonant conversion, but 
not both; 
\vspace{-0.2cm}
\item the earth matter effects must also be taken into account, since  
in $\sim 60 \,\%$ of all possible neutrino arrival times the neutrinos 
cross a substantial amount of terrestrial matter before reaching 
at least one of the existing detectors \cite{LUN01}.  
\end{itemize} 
The neutrino energy spectra measured by the various detectors (one can 
take as representatives SK, SNO and LVD) are the main tool to draw 
conclusions about neutrino oscillations, since the {\nue}, {\anue} 
and {\nux} spectra will be distorted with respect to the no mixing case, 
with a degree of distorsion strongly dependent on the type of mixing 
mechanism.

The {\nue} neutronization peak can be completely destroyed 
by an efficient conversion; this would be signalled by the 
comparison of {\nux} events in the first $10~{\rm ms}$. In this case one 
should expect a hard {\nue} spectrum in the cooling phase, coming 
from $\nu_{x} \rightarrow \nu_{e}$ oscillations. These two features 
come together and are realyzed for all solar neutrino solutions with 
normal mass hierarchy, provided that the $\left| U_{e3} \right|^{2}$ 
parameter is such that only adiabatic conversions occur. The {\anue} spectrum 
becomes in this case crucial to distinguish between the various 
solutions: if it is not affected, the SMA solution is singled out, 
while if it exhibits distorsions, with some hardening effects, 
the LMA, LOW or VO solutions are possible. A careful analysis of the 
earth matter-induced distorsions of {\anue} spectra could allow 
to perform the final distinction. 

If, on the other hand, the {\nue} neutronization peak is made by {\nue} 
and {\nux}, the {\nue} spectrum in the cooling phase has also a 
\lq\lq soft\rq\rq~(due to the uncoverted {\nue}'s) and a \lq\lq 
hard\rq\rq~(due to the converted {\nux}'s) component. These two 
components would give rise to observable distorsions, the most 
obvious of them being the presence of {\nue} events at energies 
above the expectations, i.e. a broadening of the {\nue} 
energy spectrum. Only the SMA solution is possible 
in this case if the {\anue} spectrum exhibits no distorsions.   
Note that the supernova models predict a higher energy tail 
for {\anue} than for {\nue}; the scenario described 
above could produce an inverted situation.

\begin{figure}[htb]
\begin{center}
\mbox{
	\psfig{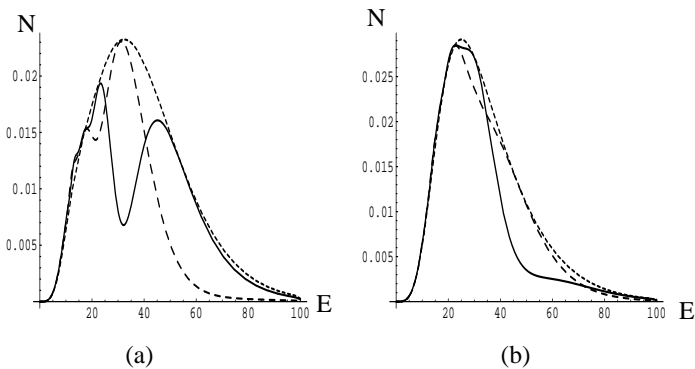}
     }
\end{center}
\caption{The earth matter effects on the {\nue} (a) and {\anue} 
(b) spectra in a scenario with LMA solution. The dotted, dashed and 
solid lines show the spectra of the normalized number of CCN 
events when the distance travelled by neutrinos through the 
earth is $0, 4000$ and $6000~{\rm Km}$ respectively (from [DIG00]).} 
\label{smirnov}
\end{figure}

If both the {\nue} and {\anue} spectra have a \lq\lq soft\rq\rq~and 
a \lq\lq hard\rq\rq~component, the situation is more complicated, 
since this scenario can be realized in many cases. However, earth-matter 
effects could help because they would be present for {\nue} and 
{\anue} in case of LMA solution, present only for {\nue} in case of SMA 
solution and completely absent in case of VO solution. Fig. \ref{smirnov} 
shows the earth-matter effects on the {\nue} and {\anue} spectra for 
some representative cases.  

Finally, a {\nue} spectrum with a \lq\lq hard\rq\rq~and a \lq\lq 
soft\rq\rq~component and a {\anue} spectrum with only the 
\lq\lq hard\rq\rq~component are obtained only with an inverted 
mass hierarchy and a $\left| U_{e3} \right|^{2}$ value which 
makes possible only adiabatic conversions, whatever is the solar 
neutrino problem solution (LMA, SMA, LOW or VO). If one of the resonant 
conversions is the true mechanism, earth effects on the {\nue} 
spectrum should be observable.  

Obviously, if the oscillation parameters are not in the selected range 
one could have different effects, for instance a pure {\nue} 
neutronization burst, which could definitely rule out the examined  
scenarios (presently the most plausible ones when all experimental data 
are taken into account). Other scenarios were also studied, 
for instance using sterile neutrinos to explain the atmospheric neutrino 
problem (this hypothesis is heavily disfavoured by SK \cite{HAB01} and 
MACRO \cite{MAC01} data) and a large ($\sim 1 \div 100~{\rm eV}$), 
cosmologically significant mass for the heavier neutrino state 
\cite{CHO00b}. The most significant variable to validate or refuse 
this scenario is the time behaviour of the ratio between charged and 
neutral current reactions (e.g. in SNO), since oscillations of higher 
energy {\nux} into {\nue} (or {\anue}) would enlarge the number of 
CCN reactions without effects on NC reactions.

\vspace{0.2cm}
The general conclusion which can be drawn is that the analysis 
of the supernova neutrino spectra could help to solve the solar 
neutrino problem (see also \cite{TAK01}), establish whether 
the mass hierarchy is normal or inverted and set more stringent 
limits on the value of $\left| U_{e3} \right|^{2}$.  

Note that, as already observed, the supernova neutrino detectors can 
not give any direct information on the {\numu}-{\nutau} mixing, 
since {\numu} and {\nutau}-induced events can not be distinguished.   

\vspace{0.2cm}
The determination of neutrino and supernova parameters discussed 
in section \ref{sec:51} would be affected by neutrino 
oscillations. In case of a strong mixing between {\nue} 
({\anue}) and {\nux} ({\anux}) one should observe a $T_{\nu_{e}}$ 
($T_{\bar{\nu}_{e}}$) of $\sim 8~{\rm MeV}$, instead of $\sim 
3 \div 4~{\rm MeV}$ in case of no mixing; on the other hand, if the mixing 
is intermediate a double peak structure should appear, produced by the 
superimposition of the \lq\lq soft\rq\rq~and \lq\lq hard\rq\rq~components. 
Since the number of CC reactions is increased by harder {\nue} or {\anue} 
spectra, the normalization (and then the determination of the source 
strength $E_{B}/D^{2}$) will also be affected. Nevertheless, the fact that 
in case of MSW resonances only one between {\nue} and {\anue} experiences 
resonant conversions would be helpful in disentangling the effects 
of neutrino oscillations.
\subsection{Combined observations of neutrino bursts and gravitational waves}
\label{sec:55}
Supernov\ae~are expected to emit not only neutrinos and light of various 
frequencies, but gravitational waves too (see e.g. \cite{SCH00}). 
The properties (amplitude, waveform, frequency ...) of the gravitational 
pulse expected from a supernova are rather uncertain, because they 
depend on the degree of non-sphericity of the collapse; the expected 
frequency range goes from $\sim 100~{\rm Hz}$ to $\sim 10~{\rm KHz}$.   
The detection of the gravitational waves is by itself of enormous 
importance, since their existence would be the most direct proof 
of the general relativity hypothesis.   

The gravitational waves are even less coupled with matter than the 
neutrinos; so, they come from the deep interior of the star and do not 
experience a slow diffusion in the supernova core. Then, a combined 
observation of neutrinos, gravitational waves and (possibly) light from 
a supernova would provide a very comprehensive picture of the 
collapse mechanism, since each of these forms of radiation preserves 
information on a different region of the stellar structure. 
Moreover, the gravitational pulse could be used to time the 
start of the collapse; this opens the possibility to set 
{\nue} mass limits at the level of fractions of {\rm eV} \cite{ARN01}
looking at the time difference between the 
{\nue} neutronization burst and the gravitational waves. 

Many gravitational wave detectors of the interferometric ({\bf VIRGO}, 
{\bf LIGO}, {\bf AIGO} ..) and resonant bar ({\bf EXPLORER}, 
{\bf NAUTILUS}, {\bf AURIGA}, {\bf ALLEGRO}, {\bf NIOBE} ..) type 
are on-line or under construction in the world; they will be sensitive 
to gravitational wave sources well beyond our galaxy. An absolute time 
resolution of a fraction of {\rm ms} (fundamental for a good 
correlation) looks within the reach of such experiments. 

The presently operating detectors, all of the bar type, are already 
sensitive to galactic supernov\ae. They form an international collaboration 
(the {\bf IGEC}, {\bf I}nternational {\bf G}ravitational {\bf E}vent 
{\bf C}ollaboration), whose aim is to produce a common analysis of 
available datasets. Note that the bar detectors have a rather narrow 
frequency band ($\Delta f \sim 50 \div 100~{\rm Hz}$), with a 
central frequency (determined by the mechanical inertia of the bar) 
of $\sim 1~{\rm KHz}$, at the center of the frequency range expected 
for a supernova. (Interferometers are instead wide-band detectors, 
sensitive to gravitational wave frequencies from few {\rm Hz} to 
tenths of {\rm KHz}.) Efforts are under way to enlarge the resonant 
bar detector bandwidth. On the other hand, the observation of an 
optical pulsar with a $T = 2.14~{\rm ms}$ emission period in SN$1987$A 
\cite{MID00} (with a modulation well explained by gravitational wave 
emission) stimulated AURIGA people to try to tune the central frequency 
of their antenna to the interesting value $f = 1/T = 467.5~{\rm Hz}$ 
\cite{BEM01}.  
\section{Conclusions}\label{sec:6}
The next galactic supernova will be a huge source of information for 
astrophysicists and particle physicists. The presence of 
a composite network of detectors, sensitive (at various levels) to 
all neutrino flavours will give the opportunity to explore many 
aspects of the collapse mechanism in large details. Correlations with 
optical and gravitational wave observations will make this collapse 
picture even more complete. Astronomers will be helped by the early  
warning provided by the SNEWS alert system and by the fast 
localization obtained by the angular distributions of the events and 
(maybe) by triangulations between various detectors.     

At the same time, the particle physicists will have a powerful source 
of neutrinos of all flavours, with exciting possibilities to set 
stringent limits on {\numu}, {\nutau} masses and on the neutrino 
oscillation mechanisms. The fact that many of the results discussed 
here are only weakly model-dependent makes this opportunity even 
more appealing.      

However, a basilar \lq\lq caveat\rq\rq~is necessary: since there are 
still many obscure points and a-priori uncertainties in the supernova 
neutrino emission, the risk of drawing rushed conclusions is just around the 
corner. Limited statistics and unknown aspects must be carefully taken 
into account when looking to supernova neutrino data.  
\section*{Acknowledgements}\label{sec:nonumber}
First of all, I want to thank all the members of the MACRO Collaboration 
for the long time spent together in working on the detector, analyze 
the data etc. Particularly, I am very grateful to my senior and younger 
Pisa colleagues (A. Baldini, C. Bemporad, M. Grassi, D. Nicol\`{o}, 
R. Pazzi and G. Signorelli) for their continuous suggestions and precious 
criticism and for reading this paper. Finally (last, but not 
least), I want to thank all the members of the Organizing Committee 
of this workshop and expecially G. Battistoni for inviting me to give 
this review talk.

\end{document}